\documentclass[longauth,traditabstract]{aa} 

\def\setsymbol#1#2{\expandafter\def\csname #1\endcsname{#2}}
\def\getsymbol#1{\csname #1\endcsname}

\def\Planck{{\it Planck\/}}

\def\HeJT{$^4$He-JT}

\def\allearlypapers{\nocite{planck2011-1.1, planck2011-1.3, planck2011-1.4, planck2011-1.5, planck2011-1.6, planck2011-1.7, planck2011-1.10, planck2011-1.10sup, planck2011-5.1a, planck2011-5.1b, planck2011-5.2a, planck2011-5.2b, planck2011-5.2c, planck2011-6.1, planck2011-6.2, planck2011-6.3a, planck2011-6.4a, planck2011-6.4b, planck2011-6.6, planck2011-7.0, planck2011-7.2, planck2011-7.3, planck2011-7.7a, planck2011-7.7b, planck2011-7.12, planck2011-7.13}}

\newbox\tablebox    \newdimen\tablewidth
\def\leaderfil{\leaders\hbox to 5pt{\hss.\hss}\hfil}
\def\endPlancktable{\tablewidth=\columnwidth 
    $$\hss\copy\tablebox\hss$$
    \vskip-\lastskip\vskip -2pt}
\def\endPlancktablewide{\tablewidth=\textwidth 
    $$\hss\copy\tablebox\hss$$
    \vskip-\lastskip\vskip -2pt}
\def\tablenote#1 #2\par{\begingroup \parindent=0.8em
    \abovedisplayshortskip=0pt\belowdisplayshortskip=0pt
    \noindent
    $$\hss\vbox{\hsize\tablewidth \hangindent=\parindent \hangafter=1 \noindent
    \hbox to \parindent{\sup{\rm #1}\hss}\strut#2\strut\par}\hss$$
    \endgroup}
\def\doubleline{\vskip 3pt\hrule \vskip 1.5pt \hrule \vskip 5pt}

\def\L2{\ifmmode L_2\else $L_2$\fi}

\def\DeltaT{\ifmmode \Delta T\else $\Delta T$\fi}
\def\deltat{\ifmmode \Delta t\else $\Delta t$\fi}
\def\fknee{\ifmmode f_{\rm knee}\else $f_{\rm knee}$\fi}
\def\Fmax{\ifmmode F_{\rm max}\else $F_{\rm max}$\fi}
\def\solar{\ifmmode{\rm M}_{\mathord\odot}\else${\rm M}_{\mathord\odot}$\fi}

\def\inv{\ifmmode^{-1}\else$^{-1}$\fi}
\def\mo{\ifmmode^{-1}\else$^{-1}$\fi}
\def\sup#1{\ifmmode ^{\rm #1}\else $^{\rm #1}$\fi}
\def\expo#1{\ifmmode \times 10^{#1}\else $\times 10^{#1}$\fi}
\def\,{\thinspace}
\def\lsim{\mathrel{\raise .4ex\hbox{\rlap{$<$}\lower 1.2ex\hbox{$\sim$}}}}
\def\gsim{\mathrel{\raise .4ex\hbox{\rlap{$>$}\lower 1.2ex\hbox{$\sim$}}}}

\def\simprop{\mathrel{\raise .4ex\hbox{\rlap{$\propto$}\lower 1.2ex\hbox{$\sim$}}}}
\def\deg{\ifmmode^\circ\else$^\circ$\fi}
\def\pdeg{\ifmmode $\setbox0=\hbox{$^{\circ}$}\rlap{\hskip.11\wd0 .}$^{\circ}
          \else \setbox0=\hbox{$^{\circ}$}\rlap{\hskip.11\wd0 .}$^{\circ}$\fi}
\def\arcs{\ifmmode {^{\scriptstyle\prime\prime}}
          \else $^{\scriptstyle\prime\prime}$\fi}
\def\arcm{\ifmmode {^{\scriptstyle\prime}}
          \else $^{\scriptstyle\prime}$\fi}
\newdimen\sa  \newdimen\sb
\def\parcs{\sa=.07em \sb=.03em
     \ifmmode \hbox{\rlap{.}}^{\scriptstyle\prime\kern -\sb\prime}\hbox{\kern -\sa}
     \else \rlap{.}$^{\scriptstyle\prime\kern -\sb\prime}$\kern -\sa\fi}
\def\parcm{\sa=.08em \sb=.03em
     \ifmmode \hbox{\rlap{.}\kern\sa}^{\scriptstyle\prime}\hbox{\kern-\sb}
     \else \rlap{.}\kern\sa$^{\scriptstyle\prime}$\kern-\sb\fi}
\def\ra[#1 #2 #3.#4]{#1\sup{h}#2\sup{m}#3\sup{s}\llap.#4}
\def\dec[#1 #2 #3.#4]{#1\deg#2\arcm#3\arcs\llap.#4}
\def\deco[#1 #2 #3]{#1\deg#2\arcm#3\arcs}
\def\rra[#1 #2]{#1\sup{h}#2\sup{m}}

\def\dots{\relax\ifmmode \ldots\else $\ldots$\fi}
\def\WHzsr{\ifmmode $W\,Hz\mo\,sr\mo$\else W\,Hz\mo\,sr\mo\fi}
\def\mHz{\ifmmode $\,mHz$\else \,mHz\fi}
\def\GHz{\ifmmode $\,GHz$\else \,GHz\fi}
\def\mKs{\ifmmode $\,mK\,s$^{1/2}\else \,mK\,s$^{1/2}$\fi}
\def\muKs{\ifmmode \,\mu$K\,s$^{1/2}\else \,$\mu$K\,s$^{1/2}$\fi}
\def\muKRJs{\ifmmode \,\mu$K$_{\rm RJ}$\,s$^{1/2}\else \,$\mu$K$_{\rm RJ}$\,s$^{1/2}$\fi}
\def\muKHz{\ifmmode \,\mu$K\,Hz$^{-1/2}\else \,$\mu$K\,Hz$^{-1/2}$\fi}
\def\MJysr{\ifmmode \,$MJy\,sr\mo$\else \,MJy\,sr\mo\fi}
\def\MJysrmK{\ifmmode \,$MJy\,sr\mo$\,mK$_{\rm CMB}\mo\else \,MJy\,sr\mo\,mK$_{\rm CMB}\mo$\fi}
\def\microns{\ifmmode \,\mu$m$\else \,$\mu$m\fi}

\def\muK{\ifmmode \,\mu$K$\else \,$\mu$\hbox{K}\fi}
\def\microK{\ifmmode \,\mu$K$\else \,$\mu$\hbox{K}\fi}
\def\muW{\ifmmode \,\mu$W$\else \,$\mu$\hbox{W}\fi}
\def\kms{\ifmmode $\,km\,s$^{-1}\else \,km\,s$^{-1}$\fi}
\def\kmsMpc{\ifmmode $\,\kms\,Mpc\mo$\else \,\kms\,Mpc\mo\fi}
\setsymbol{LFI:center:frequency:70GHz:units}{70.3\,GHz}
\setsymbol{LFI:center:frequency:44GHz:units}{44.1\,GHz}
\setsymbol{LFI:center:frequency:30GHz:units}{28.5\,GHz}

\setsymbol{LFI:center:frequency:70GHz}{70.3}
\setsymbol{LFI:center:frequency:44GHz}{44.1}
\setsymbol{LFI:center:frequency:30GHz}{28.5}

\setsymbol{LFI:center:frequency:LFI18:Rad:M:units}{71.7\GHz}
\setsymbol{LFI:center:frequency:LFI19:Rad:M:units}{67.5\GHz}
\setsymbol{LFI:center:frequency:LFI20:Rad:M:units}{69.2\GHz}
\setsymbol{LFI:center:frequency:LFI21:Rad:M:units}{70.4\GHz}
\setsymbol{LFI:center:frequency:LFI22:Rad:M:units}{71.5\GHz}
\setsymbol{LFI:center:frequency:LFI23:Rad:M:units}{70.8\GHz}
\setsymbol{LFI:center:frequency:LFI24:Rad:M:units}{44.4\GHz}
\setsymbol{LFI:center:frequency:LFI25:Rad:M:units}{44.0\GHz}
\setsymbol{LFI:center:frequency:LFI26:Rad:M:units}{43.9\GHz}
\setsymbol{LFI:center:frequency:LFI27:Rad:M:units}{28.3\GHz}
\setsymbol{LFI:center:frequency:LFI28:Rad:M:units}{28.8\GHz}
\setsymbol{LFI:center:frequency:LFI18:Rad:S:units}{70.1\GHz}
\setsymbol{LFI:center:frequency:LFI19:Rad:S:units}{69.6\GHz}
\setsymbol{LFI:center:frequency:LFI20:Rad:S:units}{69.5\GHz}
\setsymbol{LFI:center:frequency:LFI21:Rad:S:units}{69.5\GHz}
\setsymbol{LFI:center:frequency:LFI22:Rad:S:units}{72.8\GHz}
\setsymbol{LFI:center:frequency:LFI23:Rad:S:units}{71.3\GHz}
\setsymbol{LFI:center:frequency:LFI24:Rad:S:units}{44.1\GHz}
\setsymbol{LFI:center:frequency:LFI25:Rad:S:units}{44.1\GHz}
\setsymbol{LFI:center:frequency:LFI26:Rad:S:units}{44.1\GHz}
\setsymbol{LFI:center:frequency:LFI27:Rad:S:units}{28.5\GHz}
\setsymbol{LFI:center:frequency:LFI28:Rad:S:units}{28.2\GHz}

\setsymbol{LFI:center:frequency:LFI18:Rad:M}{71.7}
\setsymbol{LFI:center:frequency:LFI19:Rad:M}{67.5}
\setsymbol{LFI:center:frequency:LFI20:Rad:M}{69.2}
\setsymbol{LFI:center:frequency:LFI21:Rad:M}{70.4}
\setsymbol{LFI:center:frequency:LFI22:Rad:M}{71.5}
\setsymbol{LFI:center:frequency:LFI23:Rad:M}{70.8}
\setsymbol{LFI:center:frequency:LFI24:Rad:M}{44.4}
\setsymbol{LFI:center:frequency:LFI25:Rad:M}{44.0}
\setsymbol{LFI:center:frequency:LFI26:Rad:M}{43.9}
\setsymbol{LFI:center:frequency:LFI27:Rad:M}{28.3}
\setsymbol{LFI:center:frequency:LFI28:Rad:M}{28.8}
\setsymbol{LFI:center:frequency:LFI18:Rad:S}{70.1}
\setsymbol{LFI:center:frequency:LFI19:Rad:S}{69.6}
\setsymbol{LFI:center:frequency:LFI20:Rad:S}{69.5}
\setsymbol{LFI:center:frequency:LFI21:Rad:S}{69.5}
\setsymbol{LFI:center:frequency:LFI22:Rad:S}{72.8}
\setsymbol{LFI:center:frequency:LFI23:Rad:S}{71.3}
\setsymbol{LFI:center:frequency:LFI24:Rad:S}{44.1}
\setsymbol{LFI:center:frequency:LFI25:Rad:S}{44.1}
\setsymbol{LFI:center:frequency:LFI26:Rad:S}{44.1}
\setsymbol{LFI:center:frequency:LFI27:Rad:S}{28.5}
\setsymbol{LFI:center:frequency:LFI28:Rad:S}{28.2}

\setsymbol{LFI:white:noise:sensitivity:70GHz:units}{134.7\muKs}
\setsymbol{LFI:white:noise:sensitivity:44GHz:units}{164.7\muKs}
\setsymbol{LFI:white:noise:sensitivity:30GHz:units}{143.4\muKs}

\setsymbol{LFI:white:noise:sensitivity:70GHz}{134.7}
\setsymbol{LFI:white:noise:sensitivity:44GHz}{164.7}
\setsymbol{LFI:white:noise:sensitivity:30GHz}{143.4}

\setsymbol{LFI:white:noise:sensitivity:LFI18:Rad:M:units}{512.0\muKs}
\setsymbol{LFI:white:noise:sensitivity:LFI19:Rad:M:units}{581.4\muKs}
\setsymbol{LFI:white:noise:sensitivity:LFI20:Rad:M:units}{590.8\muKs}
\setsymbol{LFI:white:noise:sensitivity:LFI21:Rad:M:units}{455.2\muKs}
\setsymbol{LFI:white:noise:sensitivity:LFI22:Rad:M:units}{492.0\muKs}
\setsymbol{LFI:white:noise:sensitivity:LFI23:Rad:M:units}{507.7\muKs}
\setsymbol{LFI:white:noise:sensitivity:LFI24:Rad:M:units}{462.2\muKs}
\setsymbol{LFI:white:noise:sensitivity:LFI25:Rad:M:units}{413.6\muKs}
\setsymbol{LFI:white:noise:sensitivity:LFI26:Rad:M:units}{478.6\muKs}
\setsymbol{LFI:white:noise:sensitivity:LFI27:Rad:M:units}{277.7\muKs}
\setsymbol{LFI:white:noise:sensitivity:LFI28:Rad:M:units}{312.3\muKs}
\setsymbol{LFI:white:noise:sensitivity:LFI18:Rad:S:units}{465.7\muKs}
\setsymbol{LFI:white:noise:sensitivity:LFI19:Rad:S:units}{555.6\muKs}
\setsymbol{LFI:white:noise:sensitivity:LFI20:Rad:S:units}{623.2\muKs}
\setsymbol{LFI:white:noise:sensitivity:LFI21:Rad:S:units}{564.1\muKs}
\setsymbol{LFI:white:noise:sensitivity:LFI22:Rad:S:units}{534.4\muKs}
\setsymbol{LFI:white:noise:sensitivity:LFI23:Rad:S:units}{542.4\muKs}
\setsymbol{LFI:white:noise:sensitivity:LFI24:Rad:S:units}{399.2\muKs}
\setsymbol{LFI:white:noise:sensitivity:LFI25:Rad:S:units}{392.6\muKs}
\setsymbol{LFI:white:noise:sensitivity:LFI26:Rad:S:units}{418.6\muKs}
\setsymbol{LFI:white:noise:sensitivity:LFI27:Rad:S:units}{302.9\muKs}
\setsymbol{LFI:white:noise:sensitivity:LFI28:Rad:S:units}{285.3\muKs}

\setsymbol{LFI:white:noise:sensitivity:LFI18:Rad:M}{512.0}
\setsymbol{LFI:white:noise:sensitivity:LFI19:Rad:M}{581.4}
\setsymbol{LFI:white:noise:sensitivity:LFI20:Rad:M}{590.8}
\setsymbol{LFI:white:noise:sensitivity:LFI21:Rad:M}{455.2}
\setsymbol{LFI:white:noise:sensitivity:LFI22:Rad:M}{492.0}
\setsymbol{LFI:white:noise:sensitivity:LFI23:Rad:M}{507.7}
\setsymbol{LFI:white:noise:sensitivity:LFI24:Rad:M}{462.2}
\setsymbol{LFI:white:noise:sensitivity:LFI25:Rad:M}{413.6}
\setsymbol{LFI:white:noise:sensitivity:LFI26:Rad:M}{478.6}
\setsymbol{LFI:white:noise:sensitivity:LFI27:Rad:M}{277.7}
\setsymbol{LFI:white:noise:sensitivity:LFI28:Rad:M}{312.3}
\setsymbol{LFI:white:noise:sensitivity:LFI18:Rad:S}{465.7}
\setsymbol{LFI:white:noise:sensitivity:LFI19:Rad:S}{555.6}
\setsymbol{LFI:white:noise:sensitivity:LFI20:Rad:S}{623.2}
\setsymbol{LFI:white:noise:sensitivity:LFI21:Rad:S}{564.1}
\setsymbol{LFI:white:noise:sensitivity:LFI22:Rad:S}{534.4}
\setsymbol{LFI:white:noise:sensitivity:LFI23:Rad:S}{542.4}
\setsymbol{LFI:white:noise:sensitivity:LFI24:Rad:S}{399.2}
\setsymbol{LFI:white:noise:sensitivity:LFI25:Rad:S}{392.6}
\setsymbol{LFI:white:noise:sensitivity:LFI26:Rad:S}{418.6}
\setsymbol{LFI:white:noise:sensitivity:LFI27:Rad:S}{302.9}
\setsymbol{LFI:white:noise:sensitivity:LFI28:Rad:S}{285.3}

\setsymbol{LFI:knee:frequency:70GHz:units}{29.5\mHz}
\setsymbol{LFI:knee:frequency:44GHz:units}{56.2\mHz}
\setsymbol{LFI:knee:frequency:30GHz:units}{113.7\mHz}

\setsymbol{LFI:knee:frequency:70GHz}{29.5}
\setsymbol{LFI:knee:frequency:44GHz}{56.2}
\setsymbol{LFI:knee:frequency:30GHz}{113.7}

\setsymbol{LFI:knee:frequency:LFI18:Rad:M:units}{16.3\mHz}
\setsymbol{LFI:knee:frequency:LFI19:Rad:M:units}{15.1\mHz}
\setsymbol{LFI:knee:frequency:LFI20:Rad:M:units}{18.7\mHz}
\setsymbol{LFI:knee:frequency:LFI21:Rad:M:units}{37.2\mHz}
\setsymbol{LFI:knee:frequency:LFI22:Rad:M:units}{12.7\mHz}
\setsymbol{LFI:knee:frequency:LFI23:Rad:M:units}{34.6\mHz}
\setsymbol{LFI:knee:frequency:LFI24:Rad:M:units}{46.2\mHz}
\setsymbol{LFI:knee:frequency:LFI25:Rad:M:units}{24.9\mHz}
\setsymbol{LFI:knee:frequency:LFI26:Rad:M:units}{67.6\mHz}
\setsymbol{LFI:knee:frequency:LFI27:Rad:M:units}{187.4\mHz}
\setsymbol{LFI:knee:frequency:LFI28:Rad:M:units}{122.2\mHz}
\setsymbol{LFI:knee:frequency:LFI18:Rad:S:units}{17.7\mHz}
\setsymbol{LFI:knee:frequency:LFI19:Rad:S:units}{22.0\mHz}
\setsymbol{LFI:knee:frequency:LFI20:Rad:S:units}{8.7\mHz}
\setsymbol{LFI:knee:frequency:LFI21:Rad:S:units}{25.9\mHz}
\setsymbol{LFI:knee:frequency:LFI22:Rad:S:units}{15.8\mHz}
\setsymbol{LFI:knee:frequency:LFI23:Rad:S:units}{129.8\mHz}
\setsymbol{LFI:knee:frequency:LFI24:Rad:S:units}{100.9\mHz}
\setsymbol{LFI:knee:frequency:LFI25:Rad:S:units}{38.9\mHz}
\setsymbol{LFI:knee:frequency:LFI26:Rad:S:units}{58.9\mHz}
\setsymbol{LFI:knee:frequency:LFI27:Rad:S:units}{104.4\mHz}
\setsymbol{LFI:knee:frequency:LFI28:Rad:S:units}{40.7\mHz}

\setsymbol{LFI:knee:frequency:LFI18:Rad:M}{16.3}
\setsymbol{LFI:knee:frequency:LFI19:Rad:M}{15.1}
\setsymbol{LFI:knee:frequency:LFI20:Rad:M}{18.7}
\setsymbol{LFI:knee:frequency:LFI21:Rad:M}{37.2}
\setsymbol{LFI:knee:frequency:LFI22:Rad:M}{12.7}
\setsymbol{LFI:knee:frequency:LFI23:Rad:M}{34.6}
\setsymbol{LFI:knee:frequency:LFI24:Rad:M}{46.2}
\setsymbol{LFI:knee:frequency:LFI25:Rad:M}{24.9}
\setsymbol{LFI:knee:frequency:LFI26:Rad:M}{67.6}
\setsymbol{LFI:knee:frequency:LFI27:Rad:M}{187.4}
\setsymbol{LFI:knee:frequency:LFI28:Rad:M}{122.2}
\setsymbol{LFI:knee:frequency:LFI18:Rad:S}{17.7}
\setsymbol{LFI:knee:frequency:LFI19:Rad:S}{22.0}
\setsymbol{LFI:knee:frequency:LFI20:Rad:S}{8.7}
\setsymbol{LFI:knee:frequency:LFI21:Rad:S}{25.9}
\setsymbol{LFI:knee:frequency:LFI22:Rad:S}{15.8}
\setsymbol{LFI:knee:frequency:LFI23:Rad:S}{129.8}
\setsymbol{LFI:knee:frequency:LFI24:Rad:S}{100.9}
\setsymbol{LFI:knee:frequency:LFI25:Rad:S}{38.9}
\setsymbol{LFI:knee:frequency:LFI26:Rad:S}{58.9}
\setsymbol{LFI:knee:frequency:LFI27:Rad:S}{104.4}
\setsymbol{LFI:knee:frequency:LFI28:Rad:S}{40.7}

\setsymbol{LFI:slope:70GHz:units}{$-1.03$\mHz}
\setsymbol{LFI:slope:44GHz:units}{$-0.89$\mHz}
\setsymbol{LFI:slope:30GHz:units}{$-0.87$\mHz}

\setsymbol{LFI:slope:70GHz}{$-1.03$}
\setsymbol{LFI:slope:44GHz}{$-0.89$}
\setsymbol{LFI:slope:30GHz}{$-0.87$}

\setsymbol{LFI:slope:LFI18:Rad:M:units}{$-1.04$\mHz}
\setsymbol{LFI:slope:LFI19:Rad:M:units}{$-1.09$\mHz}
\setsymbol{LFI:slope:LFI20:Rad:M:units}{$-0.69$\mHz}
\setsymbol{LFI:slope:LFI21:Rad:M:units}{$-1.56$\mHz}
\setsymbol{LFI:slope:LFI22:Rad:M:units}{$-1.01$\mHz}
\setsymbol{LFI:slope:LFI23:Rad:M:units}{$-0.96$\mHz}
\setsymbol{LFI:slope:LFI24:Rad:M:units}{$-0.83$\mHz}
\setsymbol{LFI:slope:LFI25:Rad:M:units}{$-0.91$\mHz}
\setsymbol{LFI:slope:LFI26:Rad:M:units}{$-0.95$\mHz}
\setsymbol{LFI:slope:LFI27:Rad:M:units}{$-0.87$\mHz}
\setsymbol{LFI:slope:LFI28:Rad:M:units}{$-0.88$\mHz}
\setsymbol{LFI:slope:LFI18:Rad:S:units}{$-1.15$\mHz}
\setsymbol{LFI:slope:LFI19:Rad:S:units}{$-1.00$\mHz}
\setsymbol{LFI:slope:LFI20:Rad:S:units}{$-0.95$\mHz}
\setsymbol{LFI:slope:LFI21:Rad:S:units}{$-0.92$\mHz}
\setsymbol{LFI:slope:LFI22:Rad:S:units}{$-1.01$\mHz}
\setsymbol{LFI:slope:LFI23:Rad:S:units}{$-0.95$\mHz}
\setsymbol{LFI:slope:LFI24:Rad:S:units}{$-0.73$\mHz}
\setsymbol{LFI:slope:LFI25:Rad:S:units}{$-1.16$\mHz}
\setsymbol{LFI:slope:LFI26:Rad:S:units}{$-0.79$\mHz}
\setsymbol{LFI:slope:LFI27:Rad:S:units}{$-0.82$\mHz}
\setsymbol{LFI:slope:LFI28:Rad:S:units}{$-0.91$\mHz}

\setsymbol{LFI:slope:LFI18:Rad:M}{$-1.04$}
\setsymbol{LFI:slope:LFI19:Rad:M}{$-1.09$}
\setsymbol{LFI:slope:LFI20:Rad:M}{$-0.69$}
\setsymbol{LFI:slope:LFI21:Rad:M}{$-1.56$}
\setsymbol{LFI:slope:LFI22:Rad:M}{$-1.01$}
\setsymbol{LFI:slope:LFI23:Rad:M}{$-0.96$}
\setsymbol{LFI:slope:LFI24:Rad:M}{$-0.83$}
\setsymbol{LFI:slope:LFI25:Rad:M}{$-0.91$}
\setsymbol{LFI:slope:LFI26:Rad:M}{$-0.95$}
\setsymbol{LFI:slope:LFI27:Rad:M}{$-0.87$}
\setsymbol{LFI:slope:LFI28:Rad:M}{$-0.88$}
\setsymbol{LFI:slope:LFI18:Rad:S}{$-1.15$}
\setsymbol{LFI:slope:LFI19:Rad:S}{$-1.00$}
\setsymbol{LFI:slope:LFI20:Rad:S}{$-0.95$}
\setsymbol{LFI:slope:LFI21:Rad:S}{$-0.92$}
\setsymbol{LFI:slope:LFI22:Rad:S}{$-1.01$}
\setsymbol{LFI:slope:LFI23:Rad:S}{$-0.95$}
\setsymbol{LFI:slope:LFI24:Rad:S}{$-0.73$}
\setsymbol{LFI:slope:LFI25:Rad:S}{$-1.16$}
\setsymbol{LFI:slope:LFI26:Rad:S}{$-0.79$}
\setsymbol{LFI:slope:LFI27:Rad:S}{$-0.82$}
\setsymbol{LFI:slope:LFI28:Rad:S}{$-0.91$}

\setsymbol{LFI:FWHM:70GHz:units}{13\parcm01}
\setsymbol{LFI:FWHM:44GHz:units}{27\parcm92}
\setsymbol{LFI:FWHM:30GHz:units}{32\parcm65}

\setsymbol{LFI:FWHM:70GHz}{13.01}
\setsymbol{LFI:FWHM:44GHz}{27.92}
\setsymbol{LFI:FWHM:30GHz}{32.65}

\setsymbol{LFI:FWHM:LFI18:units}{13\parcm39}
\setsymbol{LFI:FWHM:LFI19:units}{13\parcm01}
\setsymbol{LFI:FWHM:LFI20:units}{12\parcm75}
\setsymbol{LFI:FWHM:LFI21:units}{12\parcm74}
\setsymbol{LFI:FWHM:LFI22:units}{12\parcm87}
\setsymbol{LFI:FWHM:LFI23:units}{13\parcm27}
\setsymbol{LFI:FWHM:LFI24:units}{22\parcm98}
\setsymbol{LFI:FWHM:LFI25:units}{30\parcm46}
\setsymbol{LFI:FWHM:LFI26:units}{30\parcm31}
\setsymbol{LFI:FWHM:LFI27:units}{32\parcm65}
\setsymbol{LFI:FWHM:LFI28:units}{32\parcm66}

\setsymbol{LFI:FWHM:LFI18}{13.39}
\setsymbol{LFI:FWHM:LFI19}{13.01}
\setsymbol{LFI:FWHM:LFI20}{12.75}
\setsymbol{LFI:FWHM:LFI21}{12.74}
\setsymbol{LFI:FWHM:LFI22}{12.87}
\setsymbol{LFI:FWHM:LFI23}{13.27}
\setsymbol{LFI:FWHM:LFI24}{22.98}
\setsymbol{LFI:FWHM:LFI25}{30.46}
\setsymbol{LFI:FWHM:LFI26}{30.31}
\setsymbol{LFI:FWHM:LFI27}{32.65}
\setsymbol{LFI:FWHM:LFI28}{32.66}

\setsymbol{LFI:FWHM:uncertainty:LFI18:units}{0.170\arcm}
\setsymbol{LFI:FWHM:uncertainty:LFI19:units}{0.174\arcm}
\setsymbol{LFI:FWHM:uncertainty:LFI20:units}{0.170\arcm}
\setsymbol{LFI:FWHM:uncertainty:LFI21:units}{0.156\arcm}
\setsymbol{LFI:FWHM:uncertainty:LFI22:units}{0.164\arcm}
\setsymbol{LFI:FWHM:uncertainty:LFI23:units}{0.171\arcm}
\setsymbol{LFI:FWHM:uncertainty:LFI24:units}{0.652\arcm}
\setsymbol{LFI:FWHM:uncertainty:LFI25:units}{1.075\arcm}
\setsymbol{LFI:FWHM:uncertainty:LFI26:units}{1.131\arcm}
\setsymbol{LFI:FWHM:uncertainty:LFI27:units}{1.266\arcm}
\setsymbol{LFI:FWHM:uncertainty:LFI28:units}{1.287\arcm}

\setsymbol{LFI:FWHM:uncertainty:LFI18}{0.170}
\setsymbol{LFI:FWHM:uncertainty:LFI19}{0.174}
\setsymbol{LFI:FWHM:uncertainty:LFI20}{0.170}
\setsymbol{LFI:FWHM:uncertainty:LFI21}{0.156}
\setsymbol{LFI:FWHM:uncertainty:LFI22}{0.164}
\setsymbol{LFI:FWHM:uncertainty:LFI23}{0.171}
\setsymbol{LFI:FWHM:uncertainty:LFI24}{0.652}
\setsymbol{LFI:FWHM:uncertainty:LFI25}{1.075}
\setsymbol{LFI:FWHM:uncertainty:LFI26}{1.131}
\setsymbol{LFI:FWHM:uncertainty:LFI27}{1.266}
\setsymbol{LFI:FWHM:uncertainty:LFI28}{1.287}

\setsymbol{HFI:center:frequency:100GHz:units}{100\,GHz}
\setsymbol{HFI:center:frequency:143GHz:units}{143\,GHz}
\setsymbol{HFI:center:frequency:217GHz:units}{217\,GHz}
\setsymbol{HFI:center:frequency:353GHz:units}{353\,GHz}
\setsymbol{HFI:center:frequency:545GHz:units}{545\,GHz}
\setsymbol{HFI:center:frequency:857GHz:units}{857\,GHz}

\setsymbol{HFI:center:frequency:100GHz}{100}
\setsymbol{HFI:center:frequency:143GHz}{143}
\setsymbol{HFI:center:frequency:217GHz}{217}
\setsymbol{HFI:center:frequency:353GHz}{353}
\setsymbol{HFI:center:frequency:545GHz}{545}
\setsymbol{HFI:center:frequency:857GHz}{857}

\setsymbol{HFI:Ndetectors:100GHz}{8}
\setsymbol{HFI:Ndetectors:143GHz}{11}
\setsymbol{HFI:Ndetectors:217GHz}{12}
\setsymbol{HFI:Ndetectors:353GHz}{12}
\setsymbol{HFI:Ndetectors:545GHz}{3}
\setsymbol{HFI:Ndetectors:857GHz}{4}

\setsymbol{HFI:FWHM:Maps:100GHz:units}{9\parcm88}
\setsymbol{HFI:FWHM:Maps:143GHz:units}{7\parcm18}
\setsymbol{HFI:FWHM:Maps:217GHz:units}{4\parcm87}
\setsymbol{HFI:FWHM:Maps:353GHz:units}{4\parcm65}
\setsymbol{HFI:FWHM:Maps:545GHz:units}{4\parcm72}
\setsymbol{HFI:FWHM:Maps:857GHz:units}{4\parcm39}
\setsymbol{HFI:FWHM:Maps:100GHz}{9.88}
\setsymbol{HFI:FWHM:Maps:143GHz}{7.18}
\setsymbol{HFI:FWHM:Maps:217GHz}{4.87}
\setsymbol{HFI:FWHM:Maps:353GHz}{4.65}
\setsymbol{HFI:FWHM:Maps:545GHz}{4.72}
\setsymbol{HFI:FWHM:Maps:857GHz}{4.39}

\setsymbol{HFI:beam:ellipticity:Maps:100GHz}{1.15}
\setsymbol{HFI:beam:ellipticity:Maps:143GHz}{1.01}
\setsymbol{HFI:beam:ellipticity:Maps:217GHz}{1.06}
\setsymbol{HFI:beam:ellipticity:Maps:353GHz}{1.05}
\setsymbol{HFI:beam:ellipticity:Maps:545GHz}{1.14}
\setsymbol{HFI:beam:ellipticity:Maps:857GHz}{1.19}

\setsymbol{HFI:FWHM:Mars:100GHz:units}{9\parcm37}
\setsymbol{HFI:FWHM:Mars:143GHz:units}{7\parcm04}
\setsymbol{HFI:FWHM:Mars:217GHz:units}{4\parcm68}
\setsymbol{HFI:FWHM:Mars:353GHz:units}{4\parcm43}
\setsymbol{HFI:FWHM:Mars:545GHz:units}{3\parcm80}
\setsymbol{HFI:FWHM:Mars:857GHz:units}{3\parcm67}

\setsymbol{HFI:FWHM:Mars:100GHz}{9.37}
\setsymbol{HFI:FWHM:Mars:143GHz}{7.04}
\setsymbol{HFI:FWHM:Mars:217GHz}{4.68}
\setsymbol{HFI:FWHM:Mars:353GHz}{4.43}
\setsymbol{HFI:FWHM:Mars:545GHz}{3.80}
\setsymbol{HFI:FWHM:Mars:857GHz}{3.67}

\setsymbol{HFI:beam:ellipticity:Mars:100GHz}{1.18}
\setsymbol{HFI:beam:ellipticity:Mars:143GHz}{1.03}
\setsymbol{HFI:beam:ellipticity:Mars:217GHz}{1.14}
\setsymbol{HFI:beam:ellipticity:Mars:353GHz}{1.09}
\setsymbol{HFI:beam:ellipticity:Mars:545GHz}{1.25}
\setsymbol{HFI:beam:ellipticity:Mars:857GHz}{1.03}

\setsymbol{HFI:CMB:relative:calibration:100GHz}{$\lsim 1\%$}
\setsymbol{HFI:CMB:relative:calibration:143GHz}{$\lsim 1\%$}
\setsymbol{HFI:CMB:relative:calibration:217GHz}{$\lsim 1\%$}
\setsymbol{HFI:CMB:relative:calibration:353GHz}{$\lsim 1\%$}
\setsymbol{HFI:CMB:relative:calibration:545GHz}{}
\setsymbol{HFI:CMB:relative:calibration:857GHz}{}

\setsymbol{HFI:CMB:absolute:calibration:100GHz}{$\lsim 2\%$}
\setsymbol{HFI:CMB:absolute:calibration:143GHz}{$\lsim 2\%$}
\setsymbol{HFI:CMB:absolute:calibration:217GHz}{$\lsim 2\%$}
\setsymbol{HFI:CMB:absolute:calibration:353GHz}{$\lsim 2\%$}
\setsymbol{HFI:CMB:absolute:calibration:545GHz}{}
\setsymbol{HFI:CMB:absolute:calibration:857GHz}{}

\setsymbol{HFI:FIRAS:gain:calibration:accuracy:statistical:100GHz}{}
\setsymbol{HFI:FIRAS:gain:calibration:accuracy:statistical:143GHz}{}
\setsymbol{HFI:FIRAS:gain:calibration:accuracy:statistical:217GHz}{}
\setsymbol{HFI:FIRAS:gain:calibration:accuracy:statistical:353GHz}{2.5\%}
\setsymbol{HFI:FIRAS:gain:calibration:accuracy:statistical:545GHz}{1\%}
\setsymbol{HFI:FIRAS:gain:calibration:accuracy:statistical:857GHz}{0.5\%}

\setsymbol{HFI:FIRAS:gain:calibration:accuracy:systematic:100GHz}{}
\setsymbol{HFI:FIRAS:gain:calibration:accuracy:systematic:143GHz}{}
\setsymbol{HFI:FIRAS:gain:calibration:accuracy:systematic:217GHz}{}
\setsymbol{HFI:FIRAS:gain:calibration:accuracy:systematic:353GHz}{}
\setsymbol{HFI:FIRAS:gain:calibration:accuracy:systematic:545GHz}{7\%}
\setsymbol{HFI:FIRAS:gain:calibration:accuracy:systematic:857GHz}{7\%}

\setsymbol{HFI:FIRAS:zero:point:accuracy:100GHz:units}{0.8\MJysr}
\setsymbol{HFI:FIRAS:zero:point:accuracy:143GHz:units}{}
\setsymbol{HFI:FIRAS:zero:point:accuracy:217GHz:units}{}
\setsymbol{HFI:FIRAS:zero:point:accuracy:353GHz:units}{1.4\MJysr}
\setsymbol{HFI:FIRAS:zero:point:accuracy:545GHz:units}{2.2\MJysr}
\setsymbol{HFI:FIRAS:zero:point:accuracy:857GHz:units}{1.7\MJysr}

\setsymbol{HFI:FIRAS:zero:point:accuracy:100GHz}{0.8}
\setsymbol{HFI:FIRAS:zero:point:accuracy:143GHz}{}
\setsymbol{HFI:FIRAS:zero:point:accuracy:217GHz}{}
\setsymbol{HFI:FIRAS:zero:point:accuracy:353GHz}{1.4}
\setsymbol{HFI:FIRAS:zero:point:accuracy:545GHz}{2.2}
\setsymbol{HFI:FIRAS:zero:point:accuracy:857GHz}{1.7}

\setsymbol{HFI:unit:conversion:100GHz:units}{0.2415\MJysrmK}
\setsymbol{HFI:unit:conversion:143GHz:units}{0.3694\MJysrmK}
\setsymbol{HFI:unit:conversion:217GHz:units}{0.4811\MJysrmK}
\setsymbol{HFI:unit:conversion:353GHz:units}{0.2883\MJysrmK}
\setsymbol{HFI:unit:conversion:545GHz:units}{0.05826\MJysrmK}
\setsymbol{HFI:unit:conversion:857GHz:units}{0.002238\MJysrmK}

\setsymbol{HFI:unit:conversion:100GHz}{0.2415}
\setsymbol{HFI:unit:conversion:143GHz}{0.3694}
\setsymbol{HFI:unit:conversion:217GHz}{0.4811}
\setsymbol{HFI:unit:conversion:353GHz}{0.2883}
\setsymbol{HFI:unit:conversion:545GHz}{0.05826}
\setsymbol{HFI:unit:conversion:857GHz}{0.002238}

\setsymbol{HFI:colour:correction:alpha=-2:V1.01:100GHz}{0.9893}
\setsymbol{HFI:colour:correction:alpha=-2:V1.01:143GHz}{0.9759}
\setsymbol{HFI:colour:correction:alpha=-2:V1.01:217GHz}{1.0007}
\setsymbol{HFI:colour:correction:alpha=-2:V1.01:353GHz}{1.0028}
\setsymbol{HFI:colour:correction:alpha=-2:V1.01:545GHz}{1.0019}
\setsymbol{HFI:colour:correction:alpha=-2:V1.01:857GHz}{0.9889}

\setsymbol{HFI:colour:correction:alpha=0:V1.01:100GHz}{1.0008}
\setsymbol{HFI:colour:correction:alpha=0:V1.01:143GHz}{1.0148}
\setsymbol{HFI:colour:correction:alpha=0:V1.01:217GHz}{0.9909}
\setsymbol{HFI:colour:correction:alpha=0:V1.01:353GHz}{0.9888}
\setsymbol{HFI:colour:correction:alpha=0:V1.01:545GHz}{0.9878}
\setsymbol{HFI:colour:correction:alpha=0:V1.01:857GHz}{1.0014}

\usepackage{natbib}
\bibpunct{(}{)}{;}{a}{}{,} %
\usepackage{graphicx}
\usepackage{txfonts}
\begin{document}

   \title{\textit{Planck} Early Results. I. The \textit{Planck} mission}

\author{\small
Planck Collaboration:
P.~A.~R.~Ade\inst{82}
\and
N.~Aghanim\inst{54}
\and
M.~Arnaud\inst{68}
\and
M.~Ashdown\inst{66, 4}
\and
J.~Aumont\inst{54}
\and
C.~Baccigalupi\inst{80}
\and
M.~Baker\inst{37}
\and
A.~Balbi\inst{31}
\and
A.~J.~Banday\inst{89, 9, 73}
\and
R.~B.~Barreiro\inst{62}
% ...
}
\institute{\small
Aalto University Mets\"{a}hovi Radio Observatory, Mets\"{a}hovintie 114, FIN-02540 Kylm\"{a}l\"{a}, Finland\\
\and
Agenzia Spaziale Italiana Science Data Center, c/o ESRIN, via Galileo Galilei, Frascati, Italy\\
\and
Astroparticule et Cosmologie, CNRS (UMR7164), Universit\'{e} Denis Diderot Paris 7, B\^{a}timent Condorcet, 10 rue A. Domon et L\'{e}onie Duquet, Paris, France\\
\and
Astrophysics Group, Cavendish Laboratory, University of Cambridge, J J Thomson Avenue, Cambridge CB3 0HE, U.K.\\
\and
Atacama Large Millimeter/submillimeter Array, ALMA Santiago Central Offices, Alonso de Cordova 3107, Vitacura, Casilla 763 0355, Santiago, Chile\\
\and
CITA, University of Toronto, 60 St. George St., Toronto, ON M5S 3H8, Canada\\
\and
CNES, 18 avenue Edouard Belin, 31401 Toulouse Cedex 9, France\\
\and
CNR - ISTI, Area della Ricerca, via G. Moruzzi 1, Pisa, Italy\\
\and
CNRS, IRAP, 9 Av. colonel Roche, BP 44346, F-31028 Toulouse cedex 4, France\\
\and
California Institute of Technology, Pasadena, California, U.S.A.\\
\and
Centre of Mathematics for Applications, University of Oslo, Blindern, Oslo, Norway\\
\and
DAMTP, University of Cambridge, Centre for Mathematical Sciences, Wilberforce Road, Cambridge CB3 0WA, U.K.\\
\and
DSM/Irfu/SPP, CEA-Saclay, F-91191 Gif-sur-Yvette Cedex, France\\
\and
DTU Space, National Space Institute, Juliane Mariesvej 30, Copenhagen, Denmark\\
\and
Departamento de F\'{\i}sica, Universidad de Oviedo, Avda. Calvo Sotelo s/n, Oviedo, Spain\\
\and
Department of Astronomy and Astrophysics, University of Toronto, 50 Saint George Street, Toronto, Ontario, Canada\\
\and
Department of Physics \& Astronomy, University of British Columbia, 6224 Agricultural Road, Vancouver, British Columbia, Canada\\
\and
Department of Physics and Astronomy, University of Southern California, Los Angeles, California, U.S.A.\\
\and
Department of Physics, Gustaf H\"{a}llstr\"{o}min katu 2a, University of Helsinki, Helsinki, Finland\\
\and
Department of Physics, Princeton University, Princeton, New Jersey, U.S.A.\\
\and
Department of Physics, Purdue University, 525 Northwestern Avenue, West Lafayette, Indiana, U.S.A.\\
\and
Department of Physics, University of California, Berkeley, California, U.S.A.\\
\and
Department of Physics, University of California, One Shields Avenue, Davis, California, U.S.A.\\
\and
Department of Physics, University of California, Santa Barbara, California, U.S.A.\\
\and
Department of Physics, University of Illinois at Urbana-Champaign, 1110 West Green Street, Urbana, Illinois, U.S.A.\\
\and
Dipartimento di Fisica G. Galilei, Universit\`{a} degli Studi di Padova, via Marzolo 8, 35131 Padova, Italy\\
\and
Dipartimento di Fisica, Universit\`{a} La Sapienza, P. le A. Moro 2, Roma, Italy\\
\and
Dipartimento di Fisica, Universit\`{a} degli Studi di Milano, Via Celoria, 16, Milano, Italy\\
\and
Dipartimento di Fisica, Universit\`{a} degli Studi di Trieste, via A. Valerio 2, Trieste, Italy\\
\and
Dipartimento di Fisica, Universit\`{a} di Ferrara, Via Saragat 1, 44122 Ferrara, Italy\\
\and
Dipartimento di Fisica, Universit\`{a} di Roma Tor Vergata, Via della Ricerca Scientifica, 1, Roma, Italy\\
\and
Discovery Center, Niels Bohr Institute, Blegdamsvej 17, Copenhagen, Denmark\\
\and
Dpto. Astrof\'{i}sica, Universidad de La Laguna (ULL), E-38206 La Laguna, Tenerife, Spain\\
\and
European Southern Observatory, ESO Vitacura, Alonso de Cordova 3107, Vitacura, Casilla 19001, Santiago, Chile\\
\and
European Space Agency, ESAC, Camino bajo del Castillo, s/n, Urbanizaci\'{o}n Villafranca del Castillo, Villanueva de la Ca\~{n}ada, Madrid, Spain\\
\and
European Space Agency, ESAC, Planck Science Office, Camino bajo del Castillo, s/n, Urbanizaci\'{o}n Villafranca del Castillo, Villanueva de la Ca\~{n}ada, Madrid, Spain\\
\and
European Space Agency, ESOC, Robert-Bosch-Str. 5, Darmstadt, Germany\\
\and
European Space Agency, ESTEC, Keplerlaan 1, 2201 AZ Noordwijk, The Netherlands\\
\and
Haverford College Astronomy Department, 370 Lancaster Avenue, Haverford, Pennsylvania, U.S.A.\\
\and
Helsinki Institute of Physics, Gustaf H\"{a}llstr\"{o}min katu 2, University of Helsinki, Helsinki, Finland\\
\and
INAF - Osservatorio Astrofisico di Catania, Via S. Sofia 78, Catania, Italy\\
\and
INAF - Osservatorio Astronomico di Padova, Vicolo dell'Osservatorio 5, Padova, Italy\\
\and
INAF - Osservatorio Astronomico di Roma, via di Frascati 33, Monte Porzio Catone, Italy\\
\and
INAF - Osservatorio Astronomico di Trieste, Via G.B. Tiepolo 11, Trieste, Italy\\
\and
INAF/IASF Bologna, Via Gobetti 101, Bologna, Italy\\
\and
INAF/IASF Milano, Via E. Bassini 15, Milano, Italy\\
\and
INRIA, Laboratoire de Recherche en Informatique, Universit\'{e} Paris-Sud 11, B\^{a}timent 490, 91405 Orsay Cedex, France\\
\and
INSU, Institut des sciences de l'univers, CNRS, 3, rue Michel-Ange, 75794 Paris Cedex 16, France\\
\and
IPAG: Institut de Plan\'{e}tologie et d'Astrophysique de Grenoble, Universit\'{e} Joseph Fourier, Grenoble 1 / CNRS-INSU, UMR 5274, Grenoble, F-38041, France\\
\and
ISDC Data Centre for Astrophysics, University of Geneva, ch. d'Ecogia 16, Versoix, Switzerland\\
\and
Imperial College London, Astrophysics group, Blackett Laboratory, Prince Consort Road, London, SW7 2AZ, U.K.\\
\and
Infrared Processing and Analysis Center, California Institute of Technology, Pasadena, CA 91125, U.S.A.\\
\and
Institut N\'{e}el, CNRS, Universit\'{e} Joseph Fourier Grenoble I, 25 rue des Martyrs, Grenoble, France\\
\and
Institut d'Astrophysique Spatiale, CNRS (UMR8617) Universit\'{e} Paris-Sud 11, B\^{a}timent 121, Orsay, France\\
\and
Institut d'Astrophysique de Paris, CNRS UMR7095, Universit\'{e} Pierre \& Marie Curie, 98 bis boulevard Arago, Paris, France\\
\and
Institut de Ci\`{e}ncies de l'Espai, CSIC/IEEC, Facultat de Ci\`{e}ncies, Campus UAB, Torre C5 par-2, Bellaterra 08193, Spain\\
\and
Institute for Space Sciences, Bucharest-Magurale, Romania\\
\and
Institute of Astronomy and Astrophysics, Academia Sinica, Taipei, Taiwan\\
\and
Institute of Astronomy, University of Cambridge, Madingley Road, Cambridge CB3 0HA, U.K.\\
\and
Institute of Theoretical Astrophysics, University of Oslo, Blindern, Oslo, Norway\\
\and
Instituto de Astrof\'{\i}sica de Canarias, C/V\'{\i}a L\'{a}ctea s/n, La Laguna, Tenerife, Spain\\
\and
Instituto de F\'{\i}sica de Cantabria (CSIC-Universidad de Cantabria), Avda. de los Castros s/n, Santander, Spain\\
\and
Istituto di Fisica del Plasma, CNR-ENEA-EURATOM Association, Via R. Cozzi 53, Milano, Italy\\
\and
Jet Propulsion Laboratory, California Institute of Technology, 4800 Oak Grove Drive, Pasadena, California, U.S.A.\\
\and
Jodrell Bank Centre for Astrophysics, Alan Turing Building, School of Physics and Astronomy, The University of Manchester, Oxford Road, Manchester, M13 9PL, U.K.\\
\and
Kavli Institute for Cosmology Cambridge, Madingley Road, Cambridge, CB3 0HA, U.K.\\
\and
LERMA, CNRS, Observatoire de Paris, 61 Avenue de l'Observatoire, Paris, France\\
\and
Laboratoire AIM, IRFU/Service d'Astrophysique - CEA/DSM - CNRS - Universit\'{e} Paris Diderot, B\^{a}t. 709, CEA-Saclay, F-91191 Gif-sur-Yvette Cedex, France\\
\and
Laboratoire Traitement et Communication de l'Information, CNRS (UMR 5141) and T\'{e}l\'{e}com ParisTech, 46 rue Barrault F-75634 Paris Cedex 13, France\\
\and
Laboratoire de Physique Subatomique et de Cosmologie, CNRS/IN2P3, Universit\'{e} Joseph Fourier Grenoble I, Institut National Polytechnique de Grenoble, 53 rue des Martyrs, 38026 Grenoble cedex, France\\
\and
Laboratoire de l'Acc\'{e}l\'{e}rateur Lin\'{e}aire, Universit\'{e} Paris-Sud 11, CNRS/IN2P3, Orsay, France\\
\and
Lawrence Berkeley National Laboratory, Berkeley, California, U.S.A.\\
\and
Max-Planck-Institut f\"{u}r Astrophysik, Karl-Schwarzschild-Str. 1, 85741 Garching, Germany\\
\and
MilliLab, VTT Technical Research Centre of Finland, Tietotie 3, Espoo, Finland\\
\and
National University of Ireland, Department of Experimental Physics, Maynooth, Co. Kildare, Ireland\\
\and
Niels Bohr Institute, Blegdamsvej 17, Copenhagen, Denmark\\
\and
Observational Cosmology, Mail Stop 367-17, California Institute of Technology, Pasadena, CA, 91125, U.S.A.\\
\and
Optical Science Laboratory, University College London, Gower Street, London, U.K.\\
\and
Rutherford Appleton Laboratory, Chilton, Didcot, U.K.\\
\and
SISSA, Astrophysics Sector, via Bonomea 265, 34136, Trieste, Italy\\
\and
SUPA, Institute for Astronomy, University of Edinburgh, Royal Observatory, Blackford Hill, Edinburgh EH9 3HJ, U.K.\\
\and
School of Physics and Astronomy, Cardiff University, Queens Buildings, The Parade, Cardiff, CF24 3AA, U.K.\\
\and
Space Research Institute (IKI), Russian Academy of Sciences, Profsoyuznaya Str, 84/32, Moscow, 117997, Russia\\
\and
Space Sciences Laboratory, University of California, Berkeley, California, U.S.A.\\
\and
Spitzer Science Center, 1200 E. California Blvd., Pasadena, California, U.S.A.\\
\and
Stanford University, Dept of Physics, Varian Physics Bldg, 382 Via Pueblo Mall, Stanford, California, U.S.A.\\
\and
Thales Alenia Space France, 100 Boulevard du Midi, Cannes la Bocca, France\\
\and
Universit\"{a}t Heidelberg, Institut f\"{u}r Theoretische Astrophysik, Albert-\"{U}berle-Str. 2, 69120, Heidelberg, Germany\\
\and
Universit\'{e} de Toulouse, UPS-OMP, IRAP, F-31028 Toulouse cedex 4, France\\
% ...
}

 \date{Received 10 January, 2011; accepted 31 May, 2011}



\abstract{The European Space Agency's \Planck\ satellite was launched on
14 May 2009, and has been surveying the sky
stably and continuously since 13 August 2009. Its performance is well in line
with expectations, and it will continue to
gather scientific data until the end of its cryogenic lifetime. We give an
overview of the history of \Planck\ in its first year of operations, and describe
some of the key performance aspects
of the satellite. This paper is part of a package submitted in conjunction
with \Planck's Early Release Compact Source Catalogue, the first 
data product based on \Planck\ to be released publicly.
The package describes the scientific performance of the
\Planck\ payload, and presents results on a variety of astrophysical
topics related to the sources included in the Catalogue, as well as selected
topics on diffuse emission. 
}



\authorrunning{Planck Collaboration}
\titlerunning{The \textit{Planck} mission}
   \maketitle

\section{Introduction}

\allearlypapers

The \Planck\ satellite\footnote{\Planck\
\emph{(http://www.esa.int/\Planck)} is a project of the European
Space Agency -- ESA -- with instruments provided by two scientific
Consortia funded by ESA member states (in particular the lead
countries: France and Italy) with contributions from NASA (USA), and
telescope reflectors provided in a collaboration between ESA and a
scientific Consortium led and funded by Denmark.} was launched on
14 May 2009, and has been surveying the sky stably and continuously
since 13 August 2009. \Planck\ carries a scientific payload consisting of an array
of 74 detectors sensitive to a range of frequencies between $\sim$\,25 and
$\sim1000\,$GHz, which scan the sky simultaneously and continuously 
with an angular resolution varying between $\sim$30 arcminutes at
the lowest frequencies and $\sim$5 arcminutes at the highest.
The array is arranged into two instruments: the detectors of the Low Frequency
Instrument \citep[LFI;][]{Bersanelli2010, planck2011-1.4}
are pseudo-correlation radiometers, covering
three bands centred at 30, 44, and $70\,$GHz; and the detectors of the High
Frequency Instrument \citep[HFI;][]{Lamarre2010, planck2011-1.5}
are bolometers, covering six bands centred at 100,
143, 217, 353, 545 and $857\,$GHz.  The design of
\Planck\ allows it to image the whole sky approximately twice per year,
with an unprecedented combination
of sensitivity, angular resolution, and frequency coverage.
The \Planck\ satellite, its payload, and its performance as predicted at the
time of launch, are described in 13 articles included in a special
issue (Volume 520) of Astronomy \& Astrophysics.

The main objective of \Planck\ is to measure the spatial anisotropies of
the temperature of the Cosmic Microwave Background (CMB), with an
accuracy set by fundamental astrophysical limits.  Its
level of performance will enable \Planck\ to extract essentially
all the information in the CMB temperature anisotropies. \Planck\
will also measure to high accuracy the polarisation of the CMB
anisotropies, which encodes not only a wealth of cosmological
information, but also provides a unique probe of the thermal history
of the Universe during the time when the first stars and galaxies
formed. In addition, the \Planck\ sky surveys will produce a wealth
of information on the properties of extragalactic sources and on the
dust and gas in our own Galaxy. The scientific objectives of
\Planck\ are described in detail in \cite{planck2005-bluebook}.

\begin{figure*}
   \centering
   \includegraphics[width=0.95\textwidth]{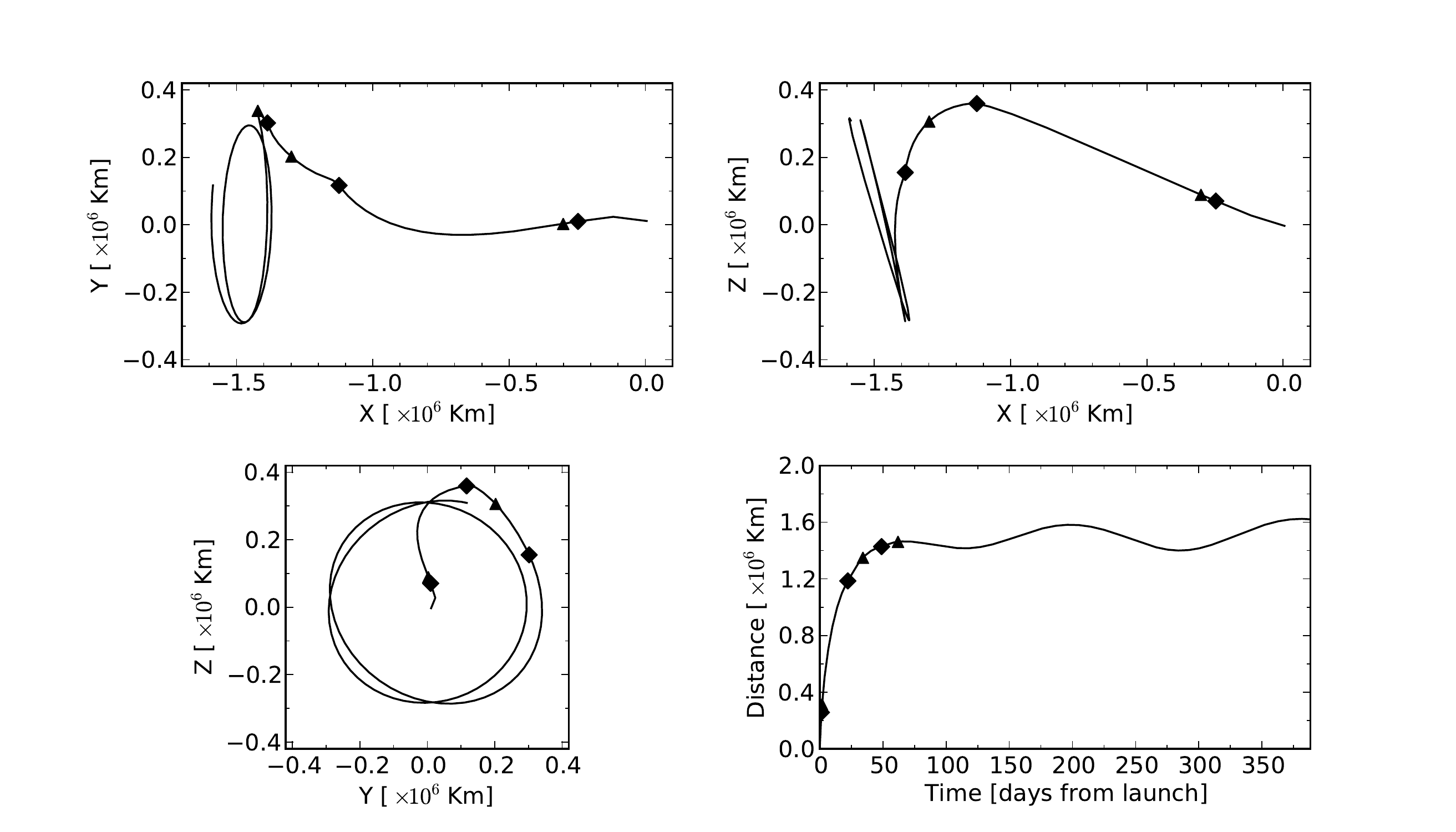}
   \caption{The trajectory of \Planck\ from launch until 6 June 2010,
in Earth-centred rotating coordinates
($X$ is in the Sun-Earth direction, and $Z$ points to the North Ecliptic Pole).
Diamond symbols indicate the major manoeuvres, while triangles
are touch-up operations.
Two orbits around L$_2$ have been carried out in this period. The orbital
periodicity is $\sim$6 months. The distance from the Earth-Moon barycentre
is shown at bottom right. }
              \label{FigOrbit}
    \end{figure*}

At the time this paper is being submitted, \Planck\ is close to
completing three surveys of the whole sky, and is releasing to the
public its first set of scientific data. This data set is the
Early Release Compact Source Catalogue (ERCSC), a list of unresolved
and compact sources extracted from the first complete all-sky
survey carried out by \Planck. The ERCSC (\cite{planck2011-1.10}) 
consists of:
\begin{itemize}
  \item nine lists of sources, extracted independently from each of \Planck's 
   nine frequency bands
  \item two lists of sources extracted using multi-band criteria
   targetted at selecting specific types of source, i.e.,
  \begin{itemize}
    \item ``Cold Cores,'' cold and dense locations in the
     Insterstellar Medium of the Milky Way, selected mainly based on their
     estimated dust temperature
    \item clusters of galaxies, selected using the spectral
     signature left on the Cosmic Microwave Background by the
     Sunyaev-Zeldovich (SZ) effect
  \end{itemize}
\end{itemize}

The ERCSC is a high-reliability compilation of sources,
released early to give the astronomical community a timely
opportunity to follow up these sources using ground- or space-based
observatories, most particularly ESA's {\it Herschel\/} 
observatory, which has a limited lifetime. The ERCSC is being released
by ESA to the public on 11 January 2011 through an online distribution
system accessible via \emph{http://www.rssd.esa.int/\Planck}.
At the same time, the \Planck\ Collaboration is submitting for publication
a package of consisting of:
\begin{itemize}
  \item this paper (\cite{planck2011-1.1}), which describes the history and
   main performance
   elements of the \Planck\ satellite in its first year of life
  \item two papers describing the performance of each of \Planck's
   two instruments (LFI and HFI) within the same period
   (\cite{planck2011-1.4} and \cite{planck2011-1.5})
  \item a paper describing the thermal performance of \Planck\ in
   orbit (\cite{planck2011-1.3})
  \item two papers describing the data processing, which has been
   applied to the data acquired by LFI and HFI, to produce the maps
   used for the ERCSC and the scientific papers in this package
   (\cite{planck2011-1.6} and \cite{planck2011-1.7})
  \item an Explanatory Supplement to the ERCSC (\cite{planck2011-1.10sup}),
   describing in detail the production and characteristics of the ERCSC
  \item a paper summarising the production of
   the ERCSC, and the main characteristics of the sources that it
   contains (\cite{planck2011-1.10})
  \item eleven papers describing in more detail: (a) specific aspects of
   different source populations contained in the ERCSC (radio
   sources, infrared galaxies, galaxy clusters, cold cores etc.);
   and (b)  cross-correlation analysis
   and follow-up observations which form part of the
   scientific validation and analysis of the ERCSC data.
These papers are: 
\begin{enumerate}
  \item \cite{planck2011-5.1a} describes the physical properties of the
   sample of clusters included in the ERCSC
  \item \cite{planck2011-5.1b} describes the validation of a subset of the
   cluster sample by follow-up observations with the {\it XMM-Newton}\ X-ray
   observatory
  \item \cite{planck2011-5.2a} analyses the statistical relationship
   between SZ flux and X-ray luminosity of the ERCSC cluster sample
  \item \cite{planck2011-5.2b} uses a high signal-to-noise subset of the
   ERCSC cluster sample to investigate the relationship between X-ray-derived
   masses and SZ fluxes
  \item \cite{planck2011-5.2c} studies the relation between SZ flux and optical
   properties of galaxy clusters by stacking \Planck\ fluxes at the locations of
   the MaxBCG optical cluster catalogue
  \item \cite{planck2011-6.1} analyses the statistical properties of a complete
   sub-sample of radio sources drawn from the ERCSC
  \item \cite{planck2011-6.2} describes the spectral energy distributions and
   other properties of some extreme radio sources, using \Planck\ ERCSC data and
   ground-based observations
  \item \cite{planck2011-6.3a} presents the spectral energy distributions of
   a sample of extragalactic radio sources, based on the \Planck\ ERCSC and
   simultaneous multi-frequency data from a range of other observatories
  \item \cite{planck2011-6.4a} studies the dust properties of nearby galaxies
   ($z<0.25$) present in the ERCSC 
  \item \cite{planck2011-7.7b} presents the statistical properties of Cold Cores 
   as observed by \Planck, in terms of spatial distribution, temperature,
   distance, mass, and morphology
  \item \cite{planck2011-7.7a} presents the physical properties and discusses
   the nature of a selection of interesting Cold Cores observed by \Planck.

\end{enumerate}
 
  \item seven papers describing in more detail selected
   science results, based on the maps which were used as input for the
   production of the ERCSC.  The results addressed in these
   papers are characterised by their robustness, a critical element
   required for publication at a rather early stage in the reduction of the
   \Planck\ data.  These seven are:
\begin{enumerate}
  \item \cite{planck2011-6.4b} presents estimates based on \Planck\ and
   {\it IRAS\/} data for the  apparent temperature and optical 
   depth of interstellar dust in the Small and Large Magellanic Clouds,
   and investigates the nature of the millimetre-wavelength excess emission
   observed in these galaxies
  \item \cite{planck2011-6.6} presents estimates of the angular power
   spectrum of the Cosmic Infrared Background as observed by \Planck\
   in selected regions of the sky
  \item \cite{planck2011-7.0} estimates over the whole sky the apparent
   temperature and optical depth of interstellar dust based on \Planck\ and
   {\it IRAS\/} data, and investigates the presence of ``dark'' gas, i.e.,
   gas which is not spatially correlated with known tracers of neutral and
   molecular gas
  \item \cite{planck2011-7.2} constructs the spectral energy distributions of
   selected regions in the Milky Way, using \Planck\ maps combined with ancillary
   multi-frequency data, and investigates the presence of anomalous excess
   emission which can be interpreted as arising from small spinning grains
  \item \cite{planck2011-7.3} estimates the radial distribution of molecular,
   neutral, and ionised gas in the Milky Way, using as spatial templates a wide
   variety of tracers of the different phases and components of the interstellar
   medium
  \item \cite{planck2011-7.12} presents a joint analysis of \Planck, {\it IRAS\/},
   and 21-cm observations of selected high-Galactic-latitude fields, and discusses
   the properties of dust in the diffuse interstellar medium close to the Sun and
   in the Galactic halo
  \item \cite{planck2011-7.13} presents \Planck\ maps of a selection of nearby
   molecular clouds, and discusses the evolution of the emitting properties of the
   dust particles embedded in them.
\end{enumerate}

\end{itemize}

The next release of \Planck\ products will take place in January 2013, and will
cover data acquired in the period up to 27 November 2010. It will include:
\begin{itemize}
\item cleaned and calibrated data timelines for each detector
\item maps in Stokes $I$, $Q$, and $U$ for each frequency band between 30
 and $353\,$GHz, and in Stokes $I$ for the two
 highest frequency bands (545 and $857\,$GHz)
\item catalogues of compact sources extracted from the frequency maps
\item maps of the main diffuse components separated from the maps, including
 the CMB
\item scientific results based on the data released 
\end{itemize}

A third release of products is foreseen after January 2014, to cover the data
acquired beyond November 2010 and the end of \Planck\ operations.

This paper is mainly dedicated to describing the history of the mission from
launch until 6 June 2010 (the coverage period of the data used to generate the
ERCSC). It
also discusses some performance aspects of the satellite which are important
for the interpretation of its scientific output. It serves therefore
as background and reference for the suite of papers described above.
In Sects.~\ref{sec:EarlyOperations} and \ref{sec:Commissioning},
we describe the main events and
activities which took place before the start of the \Planck\ surveys.
In Sect.~\ref{sec:RoutineOperations}, we describe relevant aspects of the
\Planck\ surveys, i.e., the strategy used to
scan the sky, its thermal and radiation environment, the
pointing performance of the satellite, and the flow of data in the ground segment.
Finally, in Sect.~\ref{sec:PayloadPerformance} we summarise the scientific
performance of the payload as estimated from the first year of data and with the
current set of available data processing pipelines.


\section{Early operations and transfer to orbit}\label{sec:EarlyOperations}

\Planck\ was launched from the Centre Spatial Guyanais in Kourou
(French Guyana) on 14 May 2009 at its nominal lift-off time of 13:12
UT, on an Ariane 5 ECA rocket of Arianespace\footnote{More
information on the launch facility and the launcher are available at
\emph{http://www.arianespace.com}.}. ESA's {\it Herschel\/} observatory
was launched on the same rocket. At 13:37:55 UT, {\it Herschel\/} was
released from the rocket at an altitude of $1200\,$km; 
\Planck\ followed suit at 13:40:25 UT. The separation
attitudes of both satellites were within 0\pdeg 1 of prediction.
The Ariane rocket placed \Planck\ with excellent accuracy
(semi-major axis within 1.6\%\ of prediction), on a trajectory
towards the second Lagrangian point of the Earth-Sun system
(``L$_2$'') which is drawn in Fig.~\ref{FigOrbit}. The orbit describes a
Lissajous trajectory around L$_2$ with a $\sim$6 month period that avoids
crossing the Earth penumbra for at least 4 years.

After release from the rocket, three large manoeuvres were carried
out to place \Planck\ in its intended final orbit. The first
($14.35\,{\rm m}\,{\rm s}^{-1}$), intended to correct for errors in the rocket
injection, was executed on 15 May at 20:01:05 UT, with a slight over-performance of
0.9\%\ and an error in direction of 1\pdeg 3  (a touch-up
manoeuvre was carried out on 16 May at 07:17:36 UT). The second and major
(mid-course) manoeuvre ($153.6\,{\rm m}\,{\rm s}^{-1}$) took place
between 5 and 7 June, and a touch-up ($11.8\,{\rm m}\,{\rm s}^{-1}$) was
executed on 17 June. The third and final manoeuvre ($58.8\,{\rm m}\,{\rm s}^{-1}$),
to inject \Planck\ into its final orbit,
was executed between 2 and 3 July. The total fuel consumption of
these manoeuvres, which were carried out using \Planck's coarse
($20\,$N) thrusters, was $205\,$kg. Once in its final orbit, very small
manoeuvres are required at approximately monthly intervals
($1\,{\rm m}\,{\rm s}^{-1}$ per year) to keep \Planck\ from drifting away from
its intended path around L$_2$. The attitude manoeuvres required to follow the
scanning strategy require about $2.6\,{\rm m}\,{\rm s}^{-1}$ per year. Overall,
the excellent performance of launch and orbit manoeuvres will lead to a 
large amount ($\sim160\,$kg, or $\sim$40\%\ of initial tank loading)
of fuel remaining on board at end of mission operations.

\Planck\ started cooling down radiatively shortly after launch.
Heaters were activated to hold the focal plane at $250\,$K,
which was reached around 5 hours after launch. 
The valve opening the exhaust piping of the
dilution cooler was activated at 03:30 UT, and the \HeJT\ cooler
compressors were turned on at low stroke at 05:20 UT. After these
essential operations were completed, on the second day after launch,
the focal plane temperature was allowed to descend to $170\,$K 
for out-gassing and decontamination of the
telescope and focal plane.

\section{Commissioning and Initial Science Operations}\label{sec:Commissioning}

\begin{figure*}
   \centering
   \includegraphics[width=0.9\textwidth]{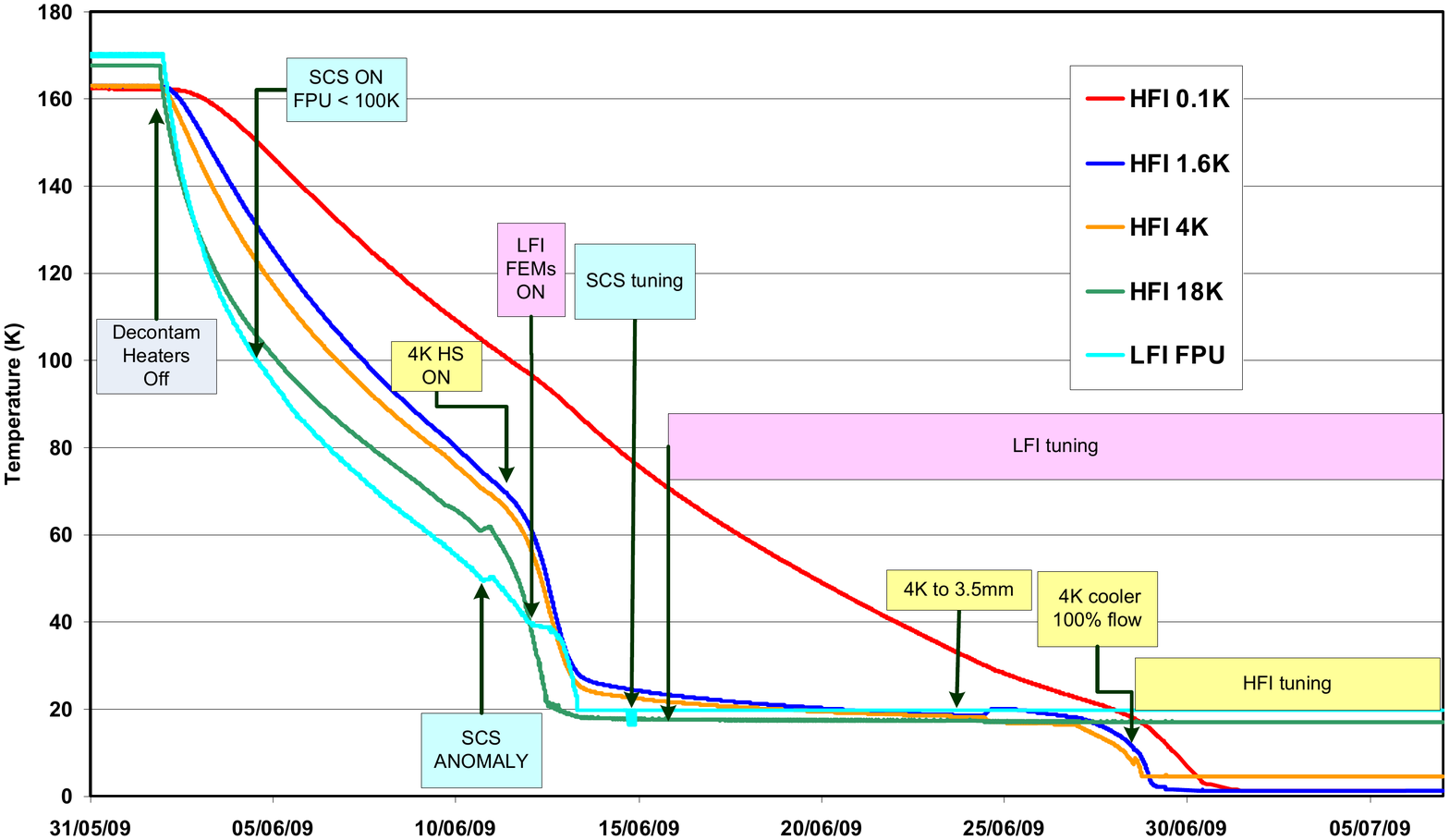}
   \caption {A sketch of the cool-down sequence, indicating when the main
   instrument and cryo-chain-related activities took place in the early phases
   of the mission. The coloured curves trace the temperature of each stage in the
   cryogenic chain.  The arrows indicate some of the key events in the sequence,
   as described in the text.  (Note: SCS = Sorption Cooler System; FEM = LFI
   Front-end Modules; FPU = Focal Plane Unit; and HS = Heat Switch). The
   events are colour-coded by sub-system: blue for the sorption cooler;
   pink for LFI; and yellow for HFI. ``LFI tuning'' refers 
   mainly to the optimisation of the bias settings of the FEMs, whereas
   ``HFI tuning'' refers mainly to optimisation of the thermal control loops at
   each low-temperature stage. }
              \label{CommissioningTimeline}
    \end{figure*}

The first period of operations focussed on commissioning activities,
i.e., functional check-out procedures of all sub-systems and
instruments of the \Planck\ spacecraft in preparation for running
science operations related to calibration and performance
verification of the payload. Planning for commissioning operations
was driven by the  telescope decontamination period of 2 weeks and
the subsequent cryogenic cool-down of the payload and instruments. The
overall duration of the cool-down was approximately 2 months,
including the decontamination
period. 

The sequence of commissioning activities covered the following
areas:
\begin{itemize}
  \item on-board commanding and data management
  \item attitude measurement and control
  \item manoeuvreing ability and orbit control
  \item telemetry and telecommand
  \item power control
  \item thermal control
  \item payload basic functionality, including:
  \begin{itemize}
    \item the LFI
    \item the HFI
    \item the cryogenic chain
    \item the Standard Radiation Environment Monitor (SREM, See
     Sect.~\ref{sec:SREM})
    \item the Fibre-Optic Gyro unit (FOG), a piggy-back experiment which is
     not used as part of the attitude control system
  \end{itemize}
\end{itemize}

The commissioning activities were executed very smoothly and all
sub-systems were found to be in good health.
Fig.~\ref{CommissioningTimeline} shows a sketch of the cool-down sequence
indicating when the main instrument-related commissioning activities
took place. The most significant unexpected issues that had to be
addressed during these early operational phases were the following.
\begin{itemize}
  \item The X-band transponder showed an initialisation anomaly during
   switch-on which was fixed by a software patch.
  \item Large reorientations of the spin axis were imperfectly completed
   and required optimisation of
   the on-board parameters of the attitude control system.
  \item The data rate required to transmit all science data to the ground
   was larger than planned, due to the unexpectedly high level of Galactic
   cosmic rays (see Sect.~\ref{sec:SREM}),
which led
  to a high glitch rate on the data stream of the HFI bolometers \citep{planck2011-1.5}; glitches increase
the dynamic range and consequently the data rate. The total data rate was controlled
by increasing the compression level of a few less critical thermometers.
  \item The level of thermal fluctuations in the 20-K stage was
   higher than originally expected. Optimisation of the sorption cooler
   operation led to an improvement, though they still remained $\sim$25\%\ higher than
   expected \citep{planck2011-1.3}.
  \item The 20-K sorption cooler turned itself off on 10 June 2009, an event
   which was traced to an incorrectly set safety threshold.
  \item A small number of sudden pressure changes were observed in the \HeJT\
   cooler during its first weeks of operation, and were most likely
   due to impurities present in the cooler gas \citep{planck2011-1.3}. The events
   disappeared after some weeks, as the impurities
   became trapped in the cooler system.
  \item The \HeJT\ cooler suffered an anomalous switch to standby mode on
   6 August 2009, following a current spike in the charge regulator unit which
   controls the current levels between the cooler electronics and the satellite
   power supply (\cite{planck2011-1.3}).  The
   cooler was restarted 20 hours after the event, and the thermal stability
   of the 100-mK stage was recovered about 47 hours later. The physical cause
   of this anomaly was not found, but the problem has not recurred.
  \item Instabilities were observed in the temperature of the \HeJT\ stage,
   which were traced to interactions with lower temperature stages, similar in
   nature to instabilities observed during ground testing \citep{planck2011-1.3}.
   They were fixed by exploring and tuning the operating points of the multiple
   stages of the cryo-system.
  \item The length of the daily telecommunications period was
   increased by from 180 to 195 minutes to improve the margin available and ensure
   completion of all daily activities\footnote{Subsequent optimisations of
   operational procedures allowed the daily contact period to be reduced
   again to 3 hours.}.
\end{itemize}

The commissioning activities were formally completed at the time
when the HFI bolometer stage reached its target temperature of $100\,$mK,
on 3 July 2009 at 01:00 UT. At this time all the critical resource
budgets (power, fuel, lifetime, etc.) were found to contain very significant 
margins with respect to the original specification.

Calibration and Performance Verification (CPV) activities started
during the cool-down period and continued until the end of August 2009.
Their objectives were to:
\begin{itemize}
  \item verify that the instruments were optimally tuned and their
   performance characterised and verified
  \item perform all tests and characterisation activities which
   could not be performed during the routine phase
  \item characterise the spacecraft and telescope characteristics of
   relevance for science\footnote{Detailed optical characterisation requires the observation of planets, 
   which first came into the field-of-view in October 2009, i.e.,
   after the start of routine operations.}
  \item estimate the lifetime of the cryogenic chain
\end{itemize}

CPV activities addressed the following areas:
\begin{itemize}
  \item tuning and characterisation of the behaviour of the cryogenic chain
  \item characterisation of the thermal behaviour of the spacecraft and
   payload
  \item for each of the two instruments: tuning; characterisation and/or
   verification of performance\footnote{In the case of LFI, an optimisation
   of the detector parameters was carried out in-flight \citep{planck2011-1.4},
   whereas for HFI, it was merely verified that the on-ground settings had not
   changed \citep{planck2011-1.5}.}, 
   calibration (including thermal, RF, noise and stability, optical response);
   and data compression properties
  \item determination of the focal plane footprint on the sky
  \item verification of scanning strategy parameters
  \item characterisation of systematic effects induced by the spacecraft
   and the telescope, including:
  \begin{itemize}
    \item dependence on solar aspect angle
    \item dependence on spin
    \item interference from the RF transmitter
    \item straylight rejection
    \item pointing performance
  \end{itemize}
\end{itemize}

\begin{figure*}
   \centering
   \includegraphics[width=0.9\textwidth]{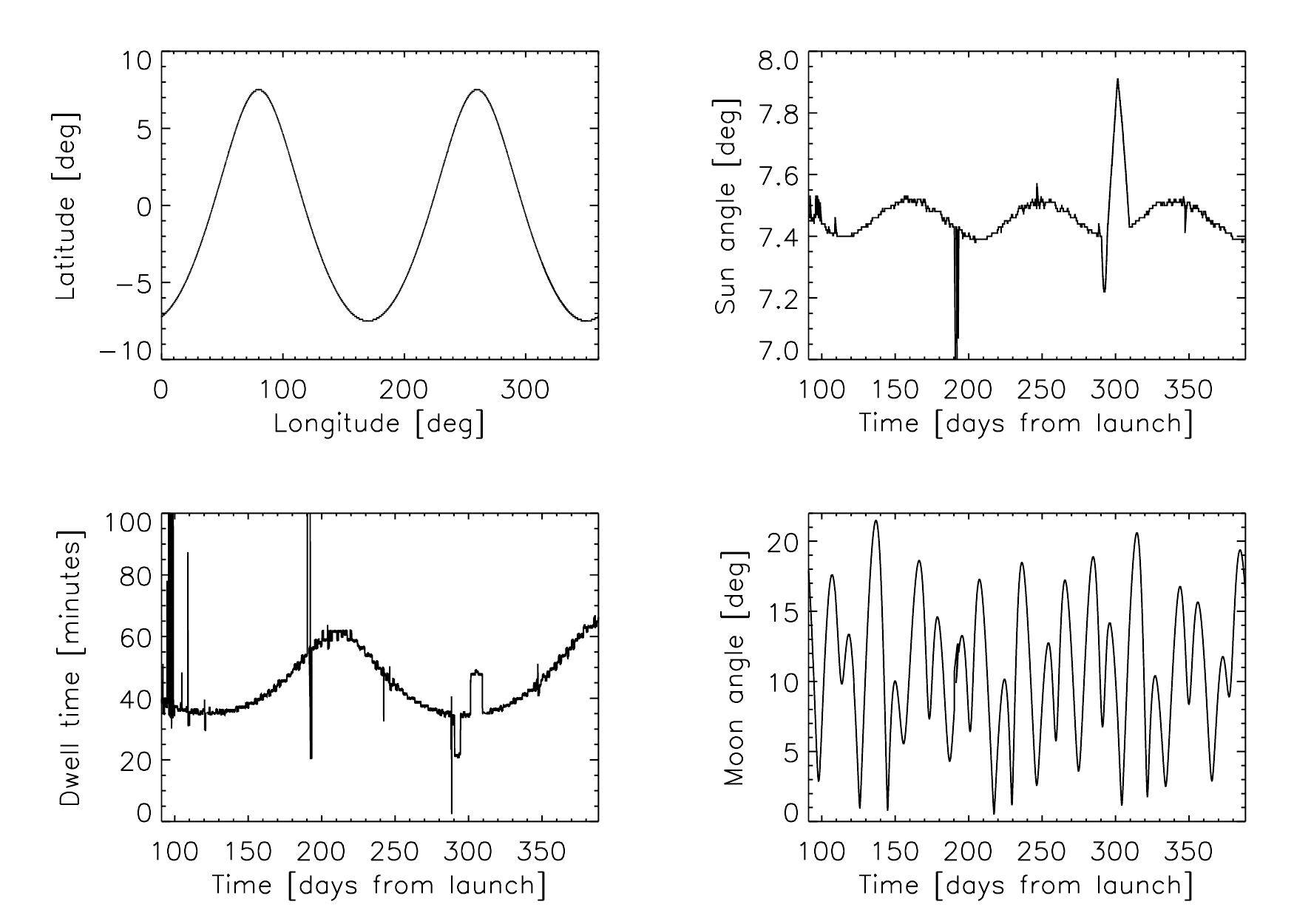} \\   
\caption{Top left: the path of the spin axis of \Planck\ (in Galactic latitude
 and longitude) over the period 13 August 2009 to 6 June 2010.
 The evolution of the dwell time (bottom left) and of the solar aspect angle
 (i.e. the angle between the anti-Sun direction and the spin axis, top right) are shown during the same period. 
 The ``day \Planck\ stood still'' (day 191) and the period of
 acceleration/deceleration during observations of the Crab (between days 291 and 310)
 are clearly visible in both plots.
 Bottom right: the evolution of the angle between the Moon and the
 anti-spin axis.
   }
              \label{FigScanning}
    \end{figure*}

The  schedule of CPV activities consumed about two weeks longer than
initially planned, mainly due to:
\begin{itemize}
  \item the anomalous switch to standby mode of the \HeJT\ cooler on 6
   August (costing 6 days until recovery)
  \item instabilities in the cryo-chain, which required the exploration of
   a larger parameter phase space to find an optimal setting point
  \item additional measurements of the voltage bias space of the LFI radiometers,
   which were introduced to optimise its noise performance, and led to the
   requirement of artificially slowing the natural cool-down of the \HeJT\ stage
\end{itemize}

A more detailed description of the relevant parts of these tests can
be found in \cite{planck2011-1.4} and \cite{planck2011-1.5}. On
completion of all the planned activities, it was concluded that:
\begin{itemize}
  \item the two instruments were fully tuned and ready for routine
   operations. No further parameter tuning was expected to be needed,
   except for the sorption cooler, which requires a weekly change in
   operational parameters \citep{planck2011-1.3}.
  \item the scientific performance parameters of both instruments
   was in most respects as had been measured on the ground before
   launch. The only significant exception was that, due to the high level
   of Galactic cosmic rays, the bolometers of HFI were
   detecting a higher number of glitches than expected, causing a 
   modest ($\sim$10\%) level of systematic effects on
   their noise properties (see details in \citep{planck2011-1.5})
  \item the telescope survived launch and cool-down in orbit without any
   major distortions or changes in its alignment
  \item the lifetime of the cryogenic chain was adequate to carry
   the mission to its foreseen end of operations in November 2010, with a
   margin of order one year
  \item the pointing performance was better than expected, and no
   changes to the planned scanning strategy were required
  \item the satellite did not introduce any major systematic effects
   into the science data. In particular, the telemetry transponder did
   not result in radio-frequency interference, which implies that the data
   acquired during visibility periods is useable for science.
\end{itemize}

The First Light Survey (FLS) was the last major activity planned
before the start of routine surveying of the sky. It was conceived as a
two-week period during which \Planck\ would be fully tuned up and
operated as if it was in its routine phase. This stable period could
have resulted in the identification of further tuning activities required
to optimise the performance of \Planck\ in the long-duration
surveys to come. The FLS was conducted between 13 and 27 August, and
in fact led to the conclusion that the \Planck\ payload was operating
stably and optimally, and required no further tuning of its
instruments. Therefore the period of the FLS was accepted as a
valid part of the first \Planck\ survey.

\section {Routine operations phase}\label{sec:RoutineOperations}

The Routine Operations phase of \Planck\ is characterised by
continuous and stable scanning of the sky and data acquisition by
LFI and HFI. It started with the FLS on 13 August of 2009, at 14:15
UT. In this section we describe the major characteristics of this
phase from start until 6 June 2010, i.e., the period over
which data were used to generate the ERCSC.

\subsection{Mission operations and data flow}

A general description of mission operations is provided in \cite{tauber2010a}.

The \Planck\ satellite generates (and stores on-board) data continuously at the
following typical rates:  $21\,{\rm kilobit}\,{\rm s}^{-1}$ (kbps) of
house-keeping (HK) data from all on-board sources, $44\,$kbps of LFI
science data and $72\,$kbps of HFI science data. The data are brought to ground
in a daily pass of approximately 3 hours duration. Besides the data downloads,
the passes also acquire real-time HK and a 20 minute period of real-time
science (used to monitor instrument performance during the pass). 
\Planck\ utilises the two ESA deep-space ground stations in New Norcia
(Australia) and Cebreros (Spain), usually the former. Scheduling of the daily
telecommunication period is quite stable, with small perturbations due to the
need to coordinate the use of the antenna with other
ESA satellites (in particular {\it Herschel\/}).

At the ground station the telemetry is received by redundant chains of
front-end/back-end equipment. The data flows to the mission operations control
centre (MOC) located at ESOC in Darmstadt (Germany), where it is processed by
redundant Mission Control Software (MCS) installations and 
made available to the science ground segment. To reduce bandwidth requirements
between the station and ESOC only one set of science telemetry is usually
transferred. Software is run post-pass to check the completeness of the data. 
This software check is also used to
build a catalogue of data completeness, which is used by the science
ground segment to control its own data transfer process. Where gaps are
detected, attempts to fill them are made as an offline activity (normally
next working day), the first step being to attempt to reflow the relevant
data from station.  Early in the mission these gaps were more frequent, with
some hundreds of packets affected per week (impact on data
return of order 50 ppm) due principally to a combination of software problems
with the data ingestion and distribution in the MCS, and imperfect behaviour
of the software gap check. Software updates implemented during the mission
have improved the situation such that gaps
are much rarer, with a total impact on data return well below 1ppm.

Redump of data from the spacecraft is attempted when there have been losses
in the space link. This has only been necessary on three occasions. 
In each case the spacecraft redump has successfully recovered all the data.

An operational principle of the mission is to avoid impact on the nominal
science of a completely missed ground station pass. Commanding continuity is
managed by keeping more than 24 hours of commanding-timeline queued on-board.
The telemetry resides on board the satellite in a $\sim$60 hour circular buffer
in solid-state memory, and can be
recovered subsequently using the margin in each pass, or more rapidly by
seeking additional station coverage after an event. The lost-pass scenario
has in fact occurred only once (on 21 December 2009), when snow on the dish at
Cebreros led to the loss of the entire pass. A rapid recovery was made by
using spare time available on the New Norcia station. Smaller impacts on the
pass occur more often (e.g., the first $\sim$10 minutes of a pass may be lost
due to a station acquisition problem) and these can normally be recovered simply
by restarting a software task
or rebooting station equipment. Such delays are normally accommodated within the
margin of the pass itself, or during the subsequent pass.

All the data downloaded from the satellite, and processed products such as
filtered attitude information, are made available each day for 
retrieval from the MOC by the LFI and HFI Data Processing Centres (DPCs).
Typically, the data arrive at the LFI (resp. HFI) DPC 2 (resp. 4) hours after the start of
the daily acquisition window. Automated
processing of the incoming telemetry is carried out each day by the LFI (resp. HFI)
DPCs and yields a daily 
data quality report which is made available to the rest of the ground segment
typically 22 (resp. 14) hours later.
More sophisticated processing of the data in each of the two DPCs is described
in \cite{planck2011-1.6}  and \cite{planck2011-1.7}.

\subsection{Scanning strategy}

\begin{figure}
   \centering
   \includegraphics[width=0.45\textwidth]{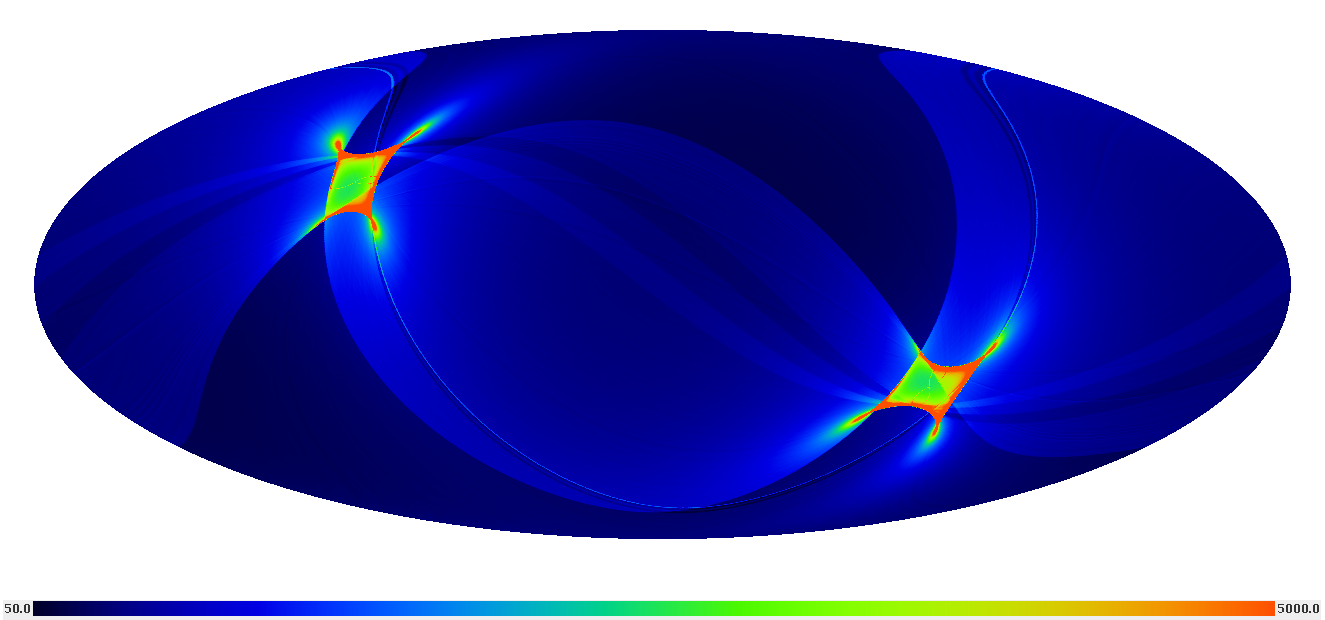} \\
   \includegraphics[width=0.45\textwidth]{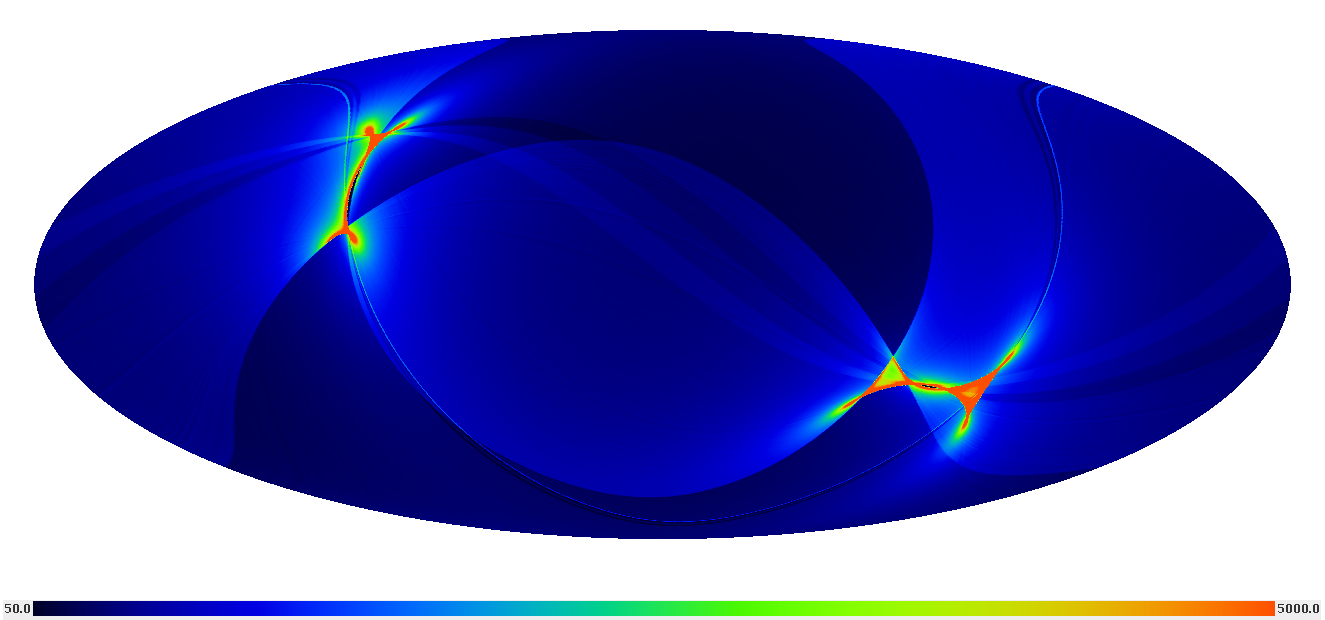} \\
   \includegraphics[width=0.45\textwidth]{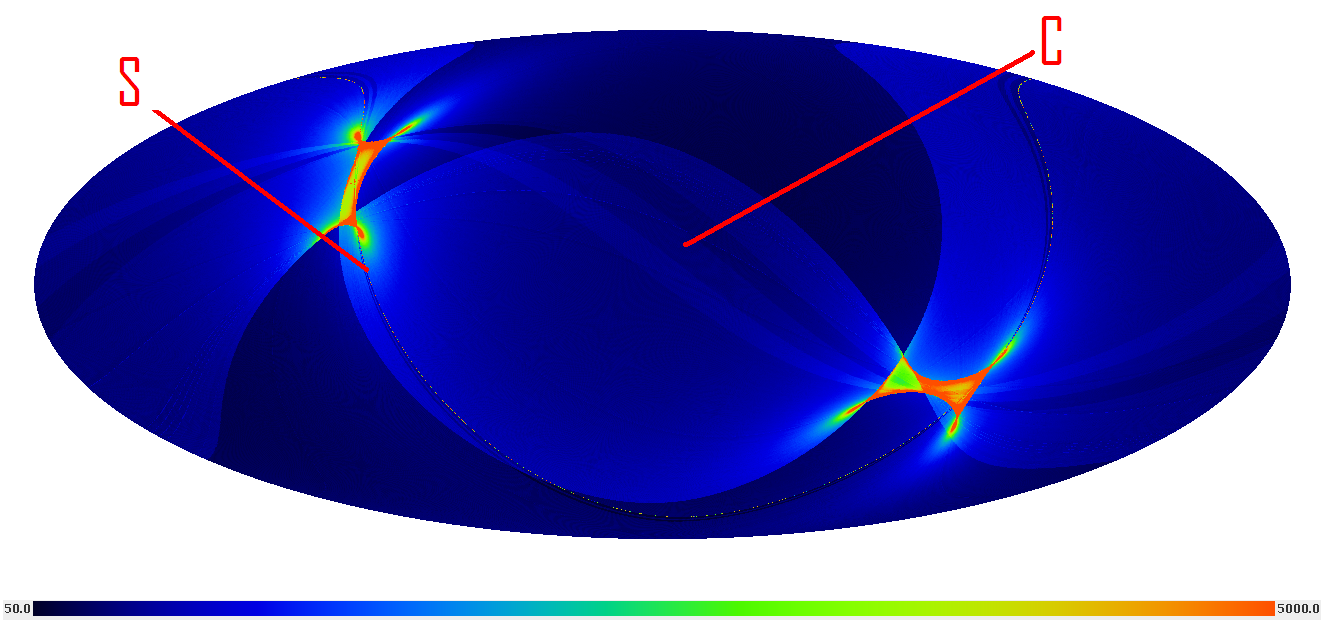} \\ \vspace{0.2cm}
   \includegraphics[width=0.35\textwidth]{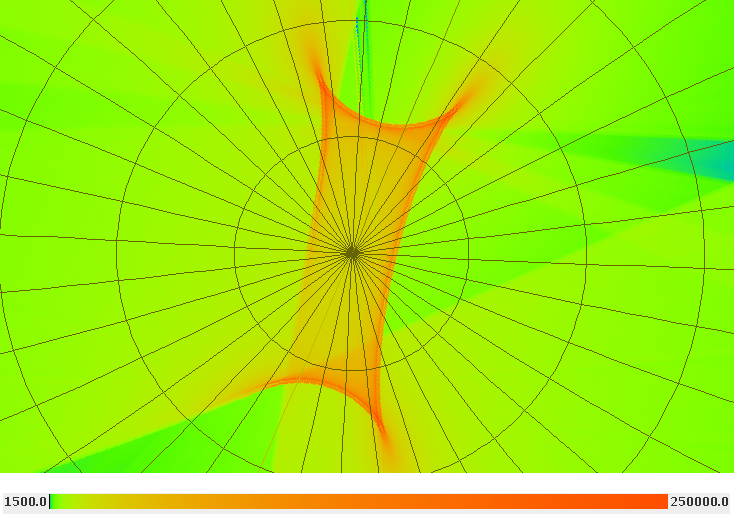} \\
\caption{Survey coverage (the colour scale represents 
 integration time varying between 50 and $5000\,{\rm sec}\,{deg}^{-2}$) for three
 individual detectors located near the edges (LFI-24 and 
 LFI-25 at $44\,$GHz, top panels) and centre of the focal plane
 (HFI 353-1 at $353\,$GHz, third panel).  The maps are Mollweide projections
 of the whole sky in Galactic coordinates, pixelised according to the Healpix
 (\cite{Gorski2005}) scheme at $N_{\rm side}$ = 1024.
The features due to ``the day Planck stood still" and the Crab slow-down
(\S\,4.2) are pointed out as ``S" and ``C" respectively. 
 The bottom panel is a zoom on the area around the North Ecliptic
 Pole, showing (in logarithmic scale) the distribution of high
 sensitivity observations integrated for all $100\,$GHz detectors.
   }
              \label{FigAllSkyCoverage}
    \end{figure}

\begin{figure}
   \centering
   \includegraphics[width=0.4\textwidth]{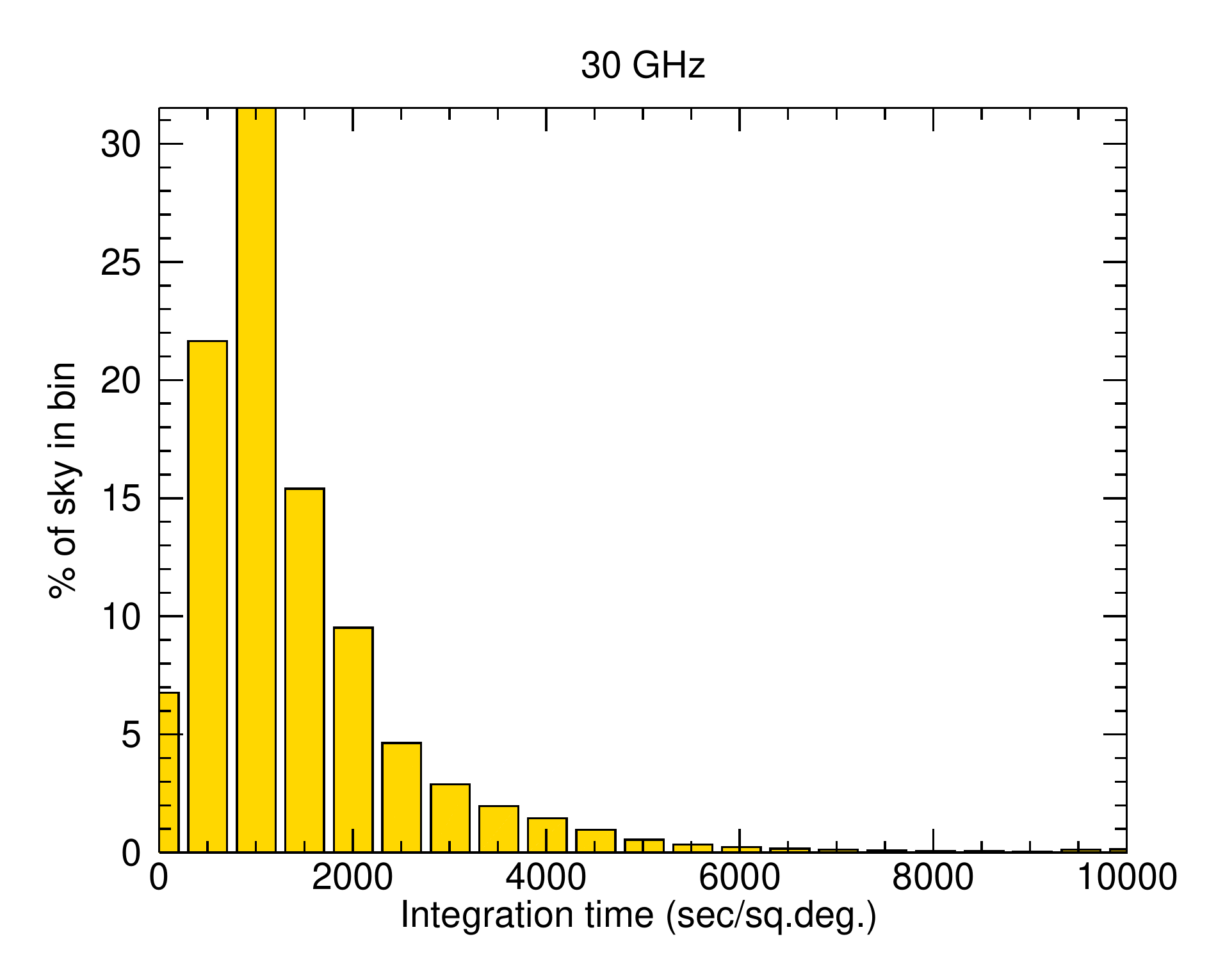} \\
   \includegraphics[width=0.4\textwidth]{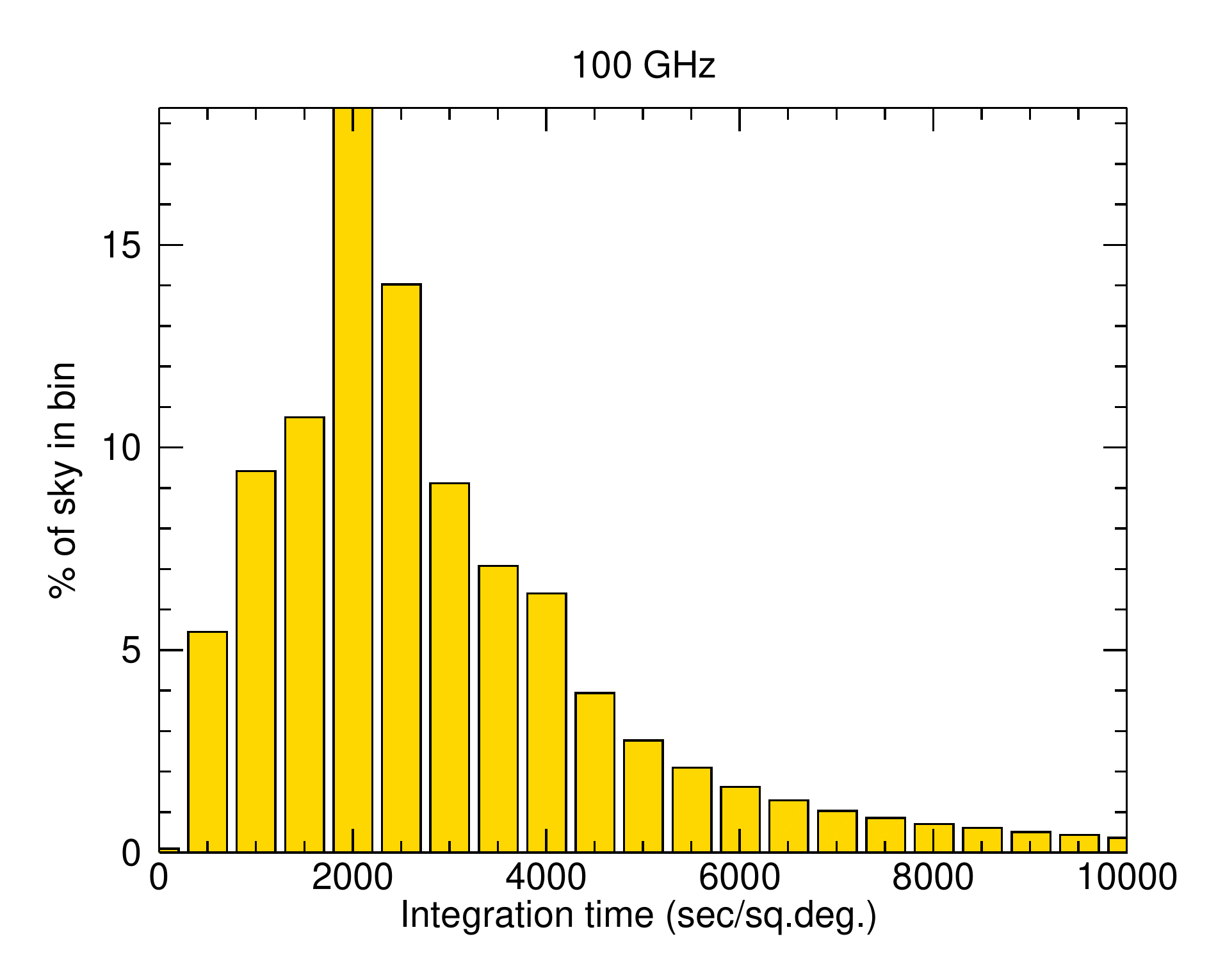} \\
   \includegraphics[width=0.4\textwidth]{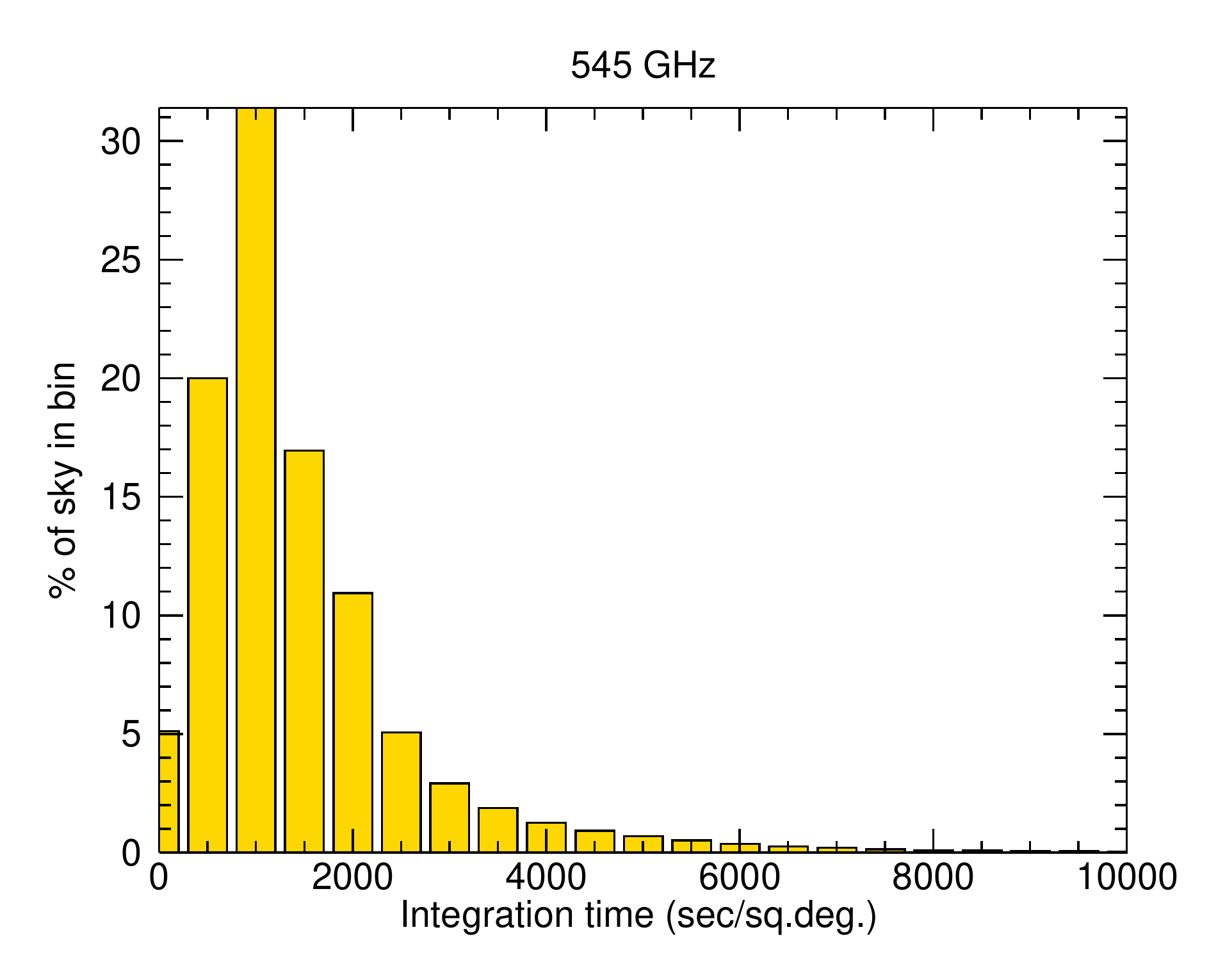} \\
   \caption{Histograms of integration time (in ${\rm s}\,{\rm deg}^{-2}$)
cumulated for all detectors at $30\,$GHz (top panel), $100\,$GHz (middle panel),
and $545\,$GHz (bottom panel). Characteristic coverage
quantities are 
listed in Table~\ref{TabCovStat}.
   }
              \label{FigHistograms}
    \end{figure}

The strategy used to scan the sky is described in \cite{tauber2010a}.
The spin axis
follows a cycloidal path on the sky as shown in Fig.~\ref{FigScanning},
by step-wise displacements of 2 arcminutes approximately every
50 minutes. The dwell time (i.e., the duration of stable data
acquisition at each pointing) has varied sinusoidally by a factor
of~$\sim$2 (see Fig.~\ref{FigScanning}). \Planck's scanning strategy
results in significantly inhomogeneous depth of integration time
across the sky;  the
areas near the ecliptic poles are observed with greater depth than
all others. This is illustrated in Fig.~\ref{FigAllSkyCoverage} and 
Fig.~\ref{FigHistograms}. Table~\ref{TabCovStat} shows more quantitatively
the coverage of the sky at three representative frequencies.

\begin{table}
\caption{\Planck\ coverage statistics.}             
\vskip -18pt
\setbox\tablebox=\vbox{
\label{TabCovStat}      
\newdimen\digitwidth
\setbox0=\hbox{\rm 0}
\digitwidth=\wd0
\catcode`*=\active
\def*{\kern\digitwidth}
\newdimen\signwidth
\setbox0=\hbox{+}
\signwidth=\wd0
\catcode`!=\active
\def!{\kern\signwidth}
\newdimen\pointwidth
\setbox0=\hbox{.}
\pointwidth=\wd0
\catcode`@=\active
\def@{\kern\pointwidth}
\halign{#\hfil\tabskip 1.0em&
\hfil#\hfil&
\hfil#\hfil&
\hfil#\hfil&
\hfil#\tabskip 8pt\cr
\noalign{\vskip 3pt\hrule\vskip 1.5pt\hrule\vskip 5pt}
& $30\,$GHz & $100\,$GHz & $545\,$GHz & \cr
\noalign{\vskip 4pt\hrule\vskip 6pt}
Mean$^{\rm a}$  & 2293@**  & 4575@**  & 2278@** & {\rm sec}\,{deg}$^{-2}$ \cr
Minimum                    & *440@**  & *801@**  & *375@** &
 {\rm sec}\,{deg}$^{-2}$ \cr
$<\,$half Mean$^{\rm b}$   & **14.4*  & **14.6*  & **15.2* & \% \cr
$>4\times$ Mean$^{\rm c}$  & ***1.6*  & ***1.5*  & ***1.2* & \% \cr
$>9\times$ Mean$^{\rm d}$  & ***0.41  & ***0.42  & ***0.41 & \% \cr
\noalign{\vskip 3pt\hrule\vskip 4pt}
} }
\endPlancktable
\tablenote a Mean over the whole sky of the integration time cumulated for
all detectors (definition as in Table~\ref{table:Instrument_performance})
in a given frequency channel.\par
\tablenote b Fraction of the sky whose coverage is less than half the Mean.\par
\tablenote c Fraction of the sky whose coverage is larger than
four times the Mean.\par
\tablenote d Fraction of the sky whose coverage is larger than
nine times the Mean.\par
\end{table}

The major pre-planned deviation from the nominal spin axis path took
place in the period 1 to 19 March 2010. During this time, the
average daily progression speed of the spin axis (normally
$1\,{\rm deg}\,{\rm day}^{-1}$)
was temporarily increased, to gain a margin with respect
to the attitude constraints imposed by the Sun and the Earth at the
time that the Crab Nebula, \Planck's main polarisation calibrator, was
being observed. This increased margin would have allowed \Planck\ to
re-observe the Crab if a significant problem had been encountered,
but none occurred. A corresponding deceleration was included to rejoin
the normal scanning path after the Crab had been observed by all
detectors. The whole operation (clearly visible in Fig.~\ref{FigScanning})
also resulted in a deviation of the solar aspect angle.

Orbit maintenance manoeuvres were carried out at
approximately monthly intervals\footnote{on 14 August 2009, 11 September 2009,
04 December 2009, 15 January 2010,
26 February 2010, and 26 March2010.}. Although the manoeuvres only required a
few minutes,
preparations, post-manoeuvre mass-property calibration, and re-entry into
scientific slewing mode increased the overhead to several
hours. The manoeuvres were carried out without disturbing the path
of the spin axis from its nominal scanning law. The dwell times of
pointings before and after the execution of the manoeuvre were
reduced to allow all pre-planned pointings to be carried out.

\begin{figure*}
   \centering
   \includegraphics[width=0.85\textwidth]{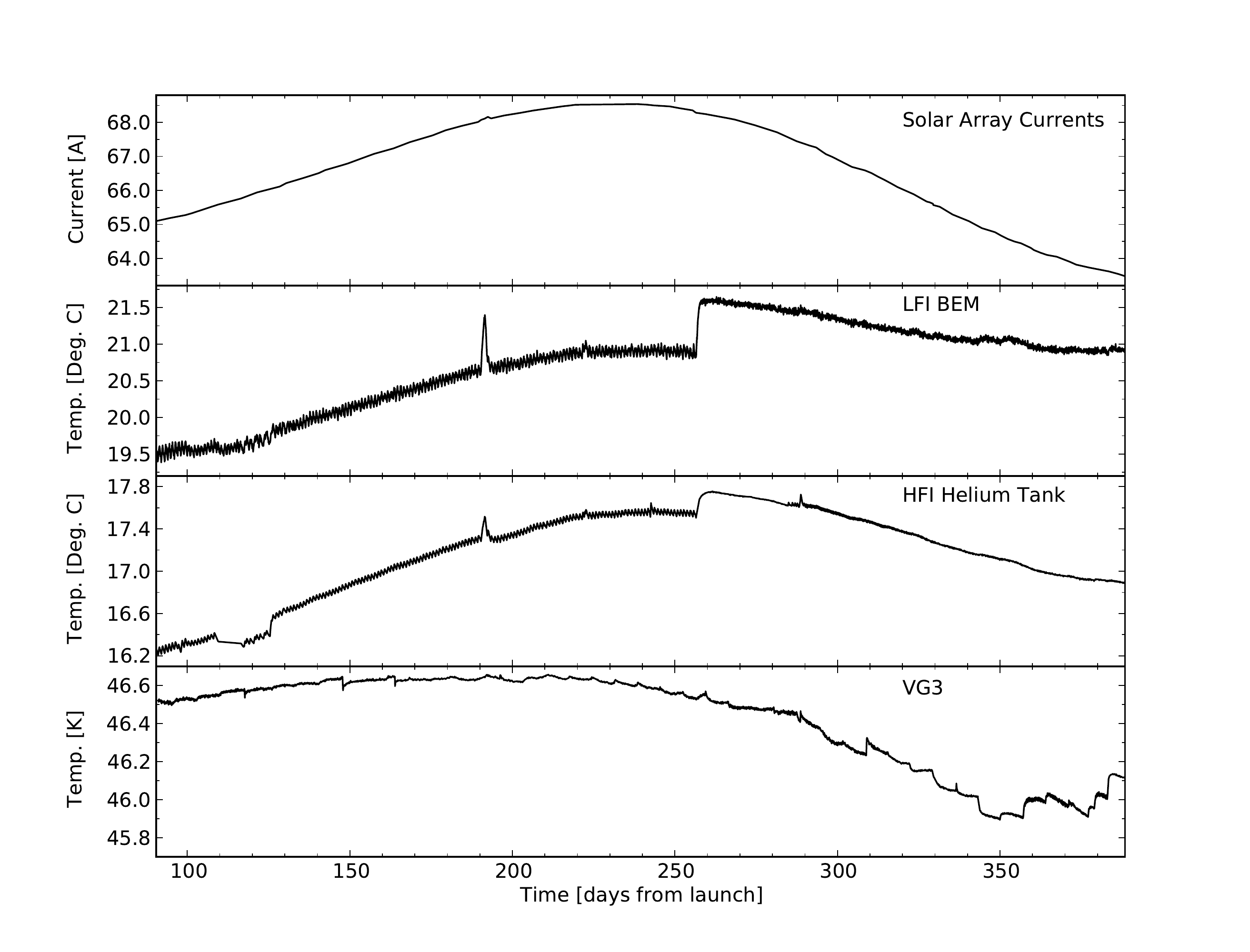} 
   \caption{
   The main long-timescale thermal modulation is a seasonal effect driven
   by the solar power absorbed by the satellite. The evolution of the solar
   heat input is traced by the top figure which shows the total current produced
   by the solar panels; the long-term variation is largely a reflection of the
   distance from the Sun, with a very small
   modulation due to variations in the satellite's aspect angle to the Sun.  
   The top and second panels show the temperature variation
   at two representative locations in the room-temperature service module (SVM),
   i.e., on one of the (HFI) Helium tanks and
   on one of the LFI back-end modules (BEM). 
The bottom panel shows the temperature evolution of VG3, the coldest of three stacked conical structures
or V-grooves which radiatively isolate the warm SVM from the cold payload module.
   The seasonal effect is not dominant in the 
evolution of VG3, demonstrating the high thermal
   isolation of the payload from the SVM.  Most variations on VG3 are due to
   weekly power input adjustments of the sorption cooler (see also
   Fig.~\ref{FigThermalEnvironment2}), which is heat-sunk to VG3.
   The main operational disturbances during the routine phase which had a
   thermal impact can be
   seen in the middle two panels (see text for more detail): (a) the ``catbed''
   event between 110 and 126 days after launch;
   (b) the ``day \Planck\ stood still'' 191 days after launch; (c)
   the change in temperature and its daily variation starting 257 days after
   launch, due to the RF transmitter being turned permanently on; and (d) two
   star-tracker reconfiguration events, 242 and  288 days after launch. 
   }
              \label{FigThermalEnvironment1}
    \end{figure*}

The main unplanned deviations from the basic scanning strategy
included the following.
\begin{itemize}
  \item An operator error in the upload of
   the on-board command timeline led to an interruption of the normal
   sequence of manoeuvres and therefore to \Planck\ pointing to the same
   location on the sky for a period of 29 hours between 20 and 21 November 2009
   (``the day \Planck\ stood still'').
   Observations of the nominal scanning pattern resumed on 22 November, and on
   23 November a recovery operation
   was applied to survey the previously missed area. During the recovery
   period the duration of pointing was decreased to allow the nominal law to
   be caught up with. As a side effect, the RF transmitter
   was left on for longer than 24 hours, which had a significant thermal
   impact on the warm part of the satellite (see Fig.~\ref{FigThermalEnvironment1}).
  \item Very minor deviations from the scanning law include occasional
   (on the average about once every two months)
   under-performance of the 1-N thrusters used for regular
   manoeuvres, which implied the corresponding pointings were
   not at the intended locations. These deviations, visible in
   Fig.~\ref{FigManError}, had typical amplitudes
   of 30\arcsec, and have no significant impact on the coverage map.
  \item During the coverage period, the operational star tracker
   switched autonomously to the redundant unit on two occasions
   (11 January 2010 and 26 February 2010);
   the nominal star tracker was restored a short period later
   (3.37 and 12.75 hours, respectively) by manual power-cycling.
   Although the science data taken during this period
   have normal quality, they have not been used because the redundant star
   tracker's performance is not fully characterised.
\end{itemize}

While the \Planck\ detectors are scanning the sky, they also
naturally observe celestial calibrators. The main objects used for
this purpose are:
\begin{itemize}
  \item the Crab Nebula, used to calibrate polarisation properties of the
   detectors, was 
   observed in September 2009, March 2010 and September 2010
  \item the brighter planets, used to map individual detector beams:
\begin{itemize}
  \item Jupiter, observed in October 2009 and July 2010
  \item Saturn, observed in January 2010 and June 2010
  \item Mars, observed in October 2009 and April 2010
\end{itemize}
\end{itemize}
The use of these observations for beam and time response calibration
is described in \cite{planck2011-1.6} and \cite{planck2011-1.7}.

The scanning strategy for the second year of Routine Operations (i.e., Surveys 3
and 4) is exactly the same as for the first year, except that all
pointings are shifted by 1 arcminute along the cross-scanning direction, in
order to provide finer sky sampling for the highest frequency detectors
when combining two years of observations.

\subsection{Thermal Environment}

\begin{figure}
   \centering
   \includegraphics[width=0.5\textwidth]{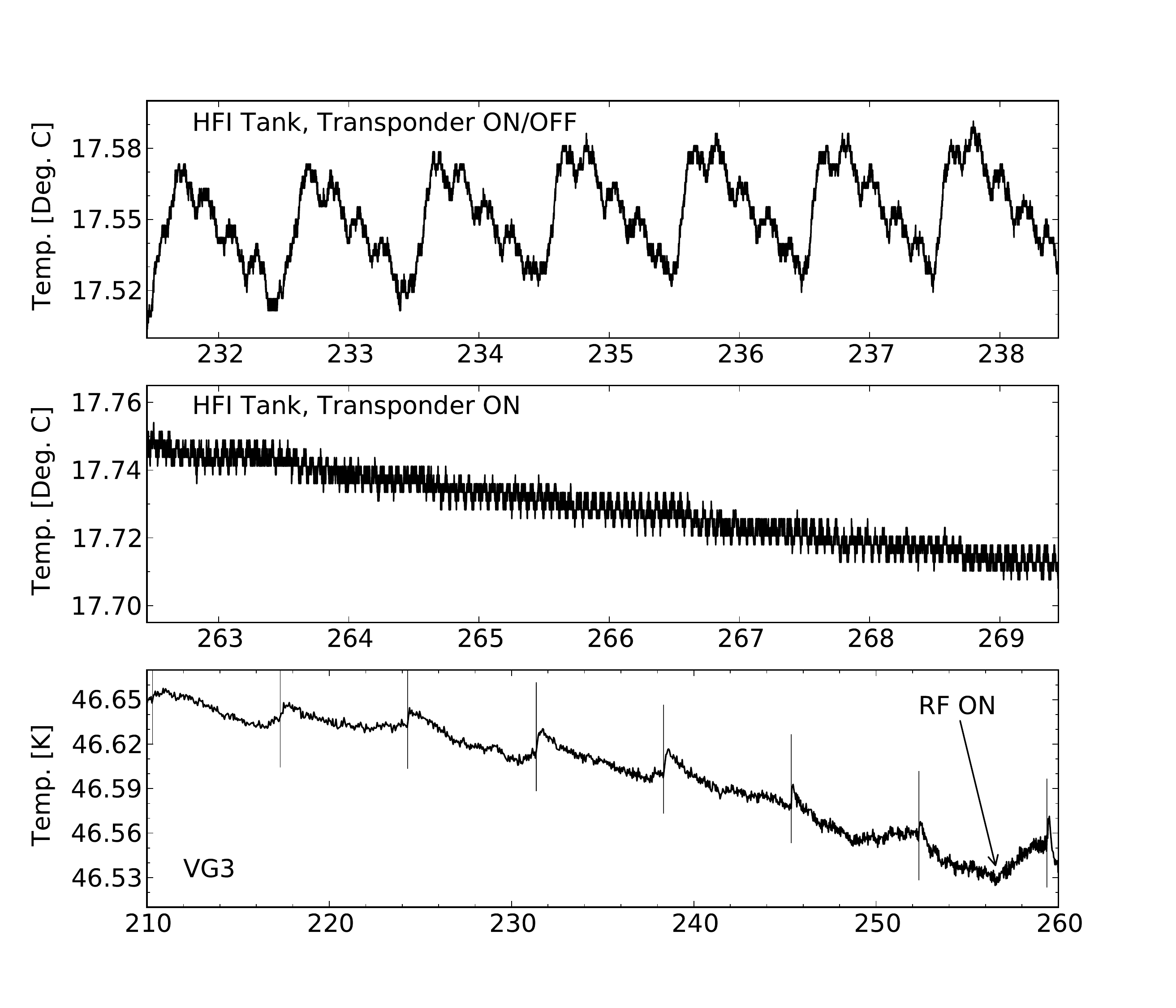}
   \caption{
 A zoom of a seven day period on the temperature of a Helium
 tank is shown before (top panel) and after (middle panel) the RF transponder
 was left permanently on (see also Fig.~\ref{FigThermalEnvironment2}), and
 illustrates the reduction in the daily
 temperature variations achieved by this operation. Smaller variations remain,
 due to thermal cycling of other elements
 (in this case the LFI Data Processing Unit), and are clearly visible in the
 middle panel. The bottom panel
 shows the typical effect of weekly updates of the operational parameters of
 the Sorption Cooler (marked as vertical lines) on the
 temperature of the third V-groove.
   }
              \label{FigThermalEnvironment2}
    \end{figure}

The satellite design and its location at L$_2$ provide an extremely
stable thermal environment (see Figs. \ref{FigThermalEnvironment1} and
\ref{FigThermalEnvironment2}).
The main temperature variation on long timescales is driven by the
total radiative power absorbed by the solar panels, which varies
depending on distance from the Sun and the solar aspect angle (i.e. the angle between the
solar direction and the spin axis).
On shorter timescales, temperature variations are driven by active thermal
regulation cycles.
Both seasonal and shorter-timescale variations are observed across the satellite's service module (SVM), but
are heavily damped and almost unobservable within the payload module (PLM).

Specific operations and deviations from the scanning strategy have a
thermal influence on the satellite and payload. Some significant effects are
clearly visible in Fig.~\ref{FigThermalEnvironment1} and listed below.
\begin{itemize}
  \item The thruster heaters
   were unintentionally turned off between 31 August and 16 September 2009
   (the so-called ``catbed'' event).
  \item As planned, the RF transmitter was initially turned on and off every day
   in synchrony with the daily visibility window, in order to reduce potential
   interference by the transmitter on the scientific data. The induced daily
   temperature variation had a measurable effect throughout the
   satellite. An important effect was on the temperature of the \HeJT\ 
   cooler compressors, which caused variations of the levels of the
   interference lines that they induce on the bolometer data
   (\cite{planck2011-1.5}). Therefore the RF transmitter was left
   permanently on starting from 25 January 2010 (257 days after launch), which made a
   noticeable improvement on the daily temperature variations
   (Fig.~\ref{FigThermalEnvironment2}).
  \item A significant thermal effect arises from the (approximately) weekly
   adjustments to the operation of the Sorption Cooler
   (Fig.~\ref{FigThermalEnvironment2}).
\end{itemize}

The thermal environment of the payload module is -- by design -- extremely well
decoupled from that of the service module (Fig.~\ref{FigThermalEnvironment1}).
As a consequence, in spite of the significant thermal perturbations originating
in the SVM, the thermal variability affecting the detectors is essentially
completely due to the operation of the cryogenic cooling chain (described in
detail in \cite{planck2011-1.3}), which ensures their cold environment.

\subsection{Radiation Environment}\label{sec:SREM}

The Standard Radiation Environment Monitor on board \Planck\ (SREM,
\cite{Buehler1996}) is a particle detector 
which is being flown on several ESA satellites. The SREM
consists of several detectors sensitive to different energy ranges,
which can also be used in coincidence mode. In particular, the SREM
measures count rates of high energy protons (from $\sim10\,$MeV to $\sim300\,$MeV)
and electrons ($\sim300\,$keV to $\sim6\,$MeV).

\begin{figure*}
   \centering
   \includegraphics[width=0.6\textwidth]{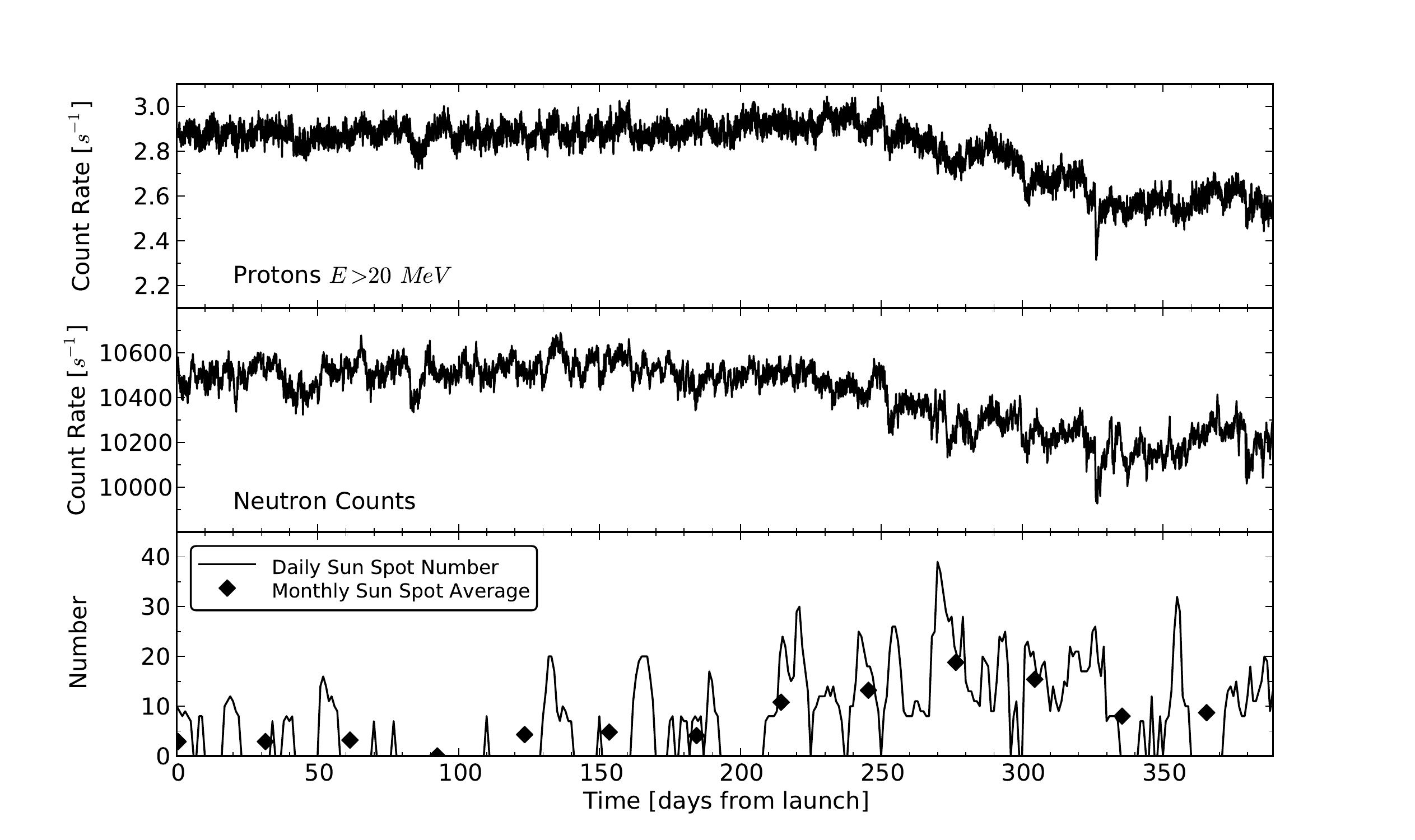} \  
   \includegraphics[width=0.35\textwidth]{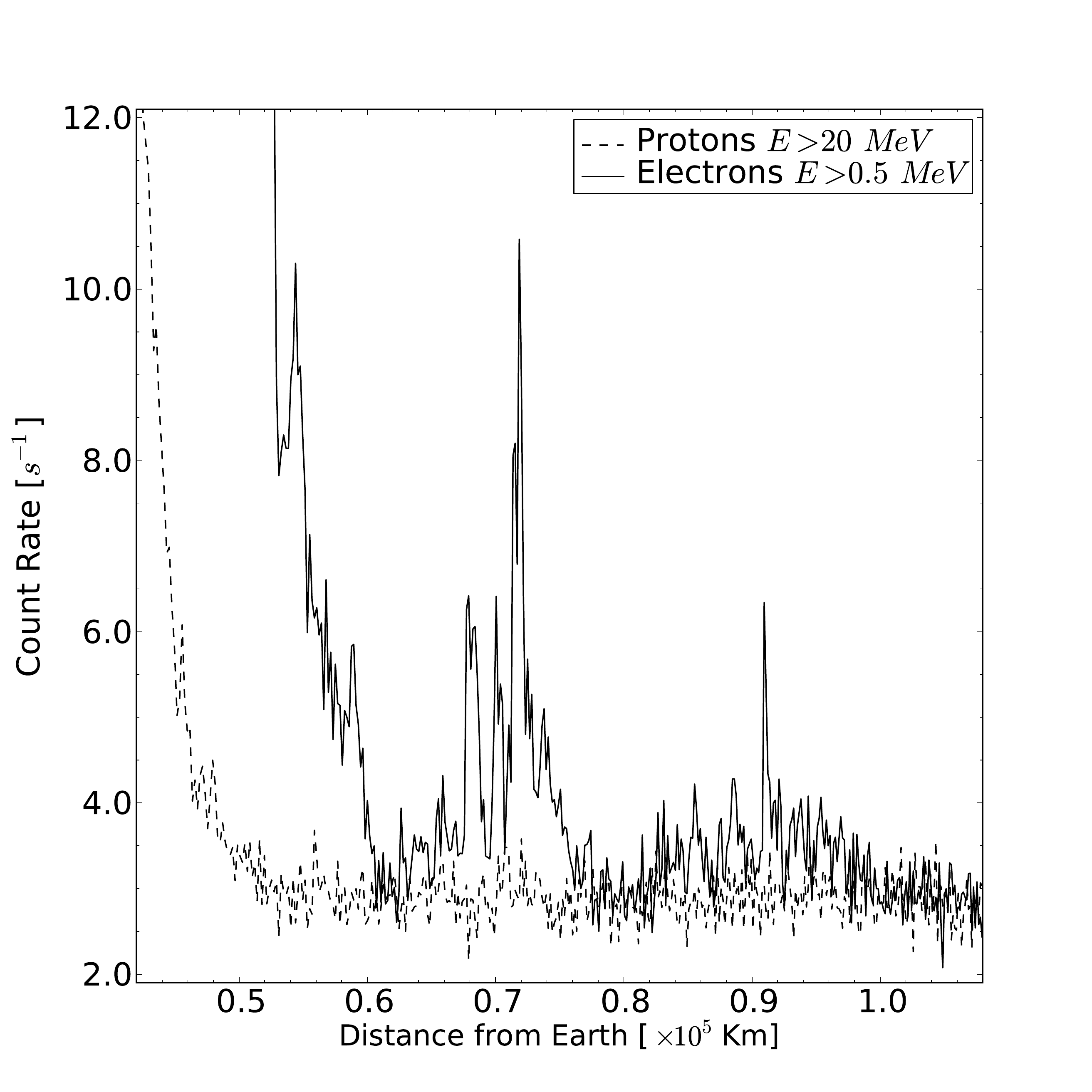} 
   \caption{The top left panel shows the time evolution of an SREM measure (TC1, \cite{Buehler1996}) which is
    sensitive to high-energy protons.  The levels shortly after launch are also
    indicated (on the right), showing the passage through
    the van Allen radiation belts, characterised by very high count levels.
    The bottom panel (on the left) shows the corresponding evolution of the
    sunspot number, indicating the slow transition of the solar
    cycle out of its current minimum. A monitor of high-energy neutrons
    (in this case located at McMurdo station, data courtesy of Bartol Research
    Institute, supported by US NSF), presented in the middle panel, traces the
    corresponding decrease in the flux of Galactic cosmic rays as the Sun
    becomes more active.}
              \label{FigRadiationEnvironment}
    \end{figure*}

Particle fluxes measured by the SREM on board \Planck\ are shown in
Fig.~\ref{FigRadiationEnvironment}. The radiation environment of
\Planck\ is characterised by the current epoch near the minimum in the solar
cycle.  As a consequence, the particle
flux is dominated by Galactic cosmic rays, rather than by the solar
wind. The time evolution of the SREM measurements is well correlated with
that of identical units flying simultaneously on other satellites (e.g.,
{\it Herschel, Rosetta\/}) and with indicators of Galactic cosmic rays, and is
anti-correlated with solar flare events and with the solar cycle
(Fig.~\ref{FigRadiationEnvironment}). More importantly
for \Planck, the SREM measurements are very well correlated with the heat
deposition on the coldest stages of the HFI, and with glitch rates
measured by the detectors of HFI. A more detailed interpretation of
these data is provided in \cite{planck2011-1.5}.

\subsection{Pointing performance}

Redundant star trackers on board \Planck\ (co-aligned within 
0\pdeg 2 with the instrument field-of-view) provide absolute attitude
measurements at a frequency of $8\,$Hz. These measurements are used by
the attitude control computer on board \Planck\ to execute the commanded
reorientations of the satellite. The star tracker data are further
processed on the ground on a daily basis, to provide:
\begin{itemize}
  \item filtered attitude information at a frequency of $4\,$Hz during the stable
   observation periods. The filtering
   algorithm basically suppresses high-frequency components of the
   measurement noise (i.e., at frequencies well above the nutation frequency)
  \item reconstructed attitude parameters averaged over each
   spin period (60 seconds) and each stable observation period (or
   dwell, typically of 50 minutes length)
\end{itemize}

The daily filtered attitude information is used by the data
processing centres to estimate the location of each detector beam
with respect to the satellite reference frame, based mainly on
observations of planets (as described in \cite{planck2011-1.5}). 
The attitude data during the periods that the
satellite is slewing are not filtered on the ground and are therefore
much noisier.

\begin{figure}
   \centering
   \includegraphics[width=0.5\textwidth]{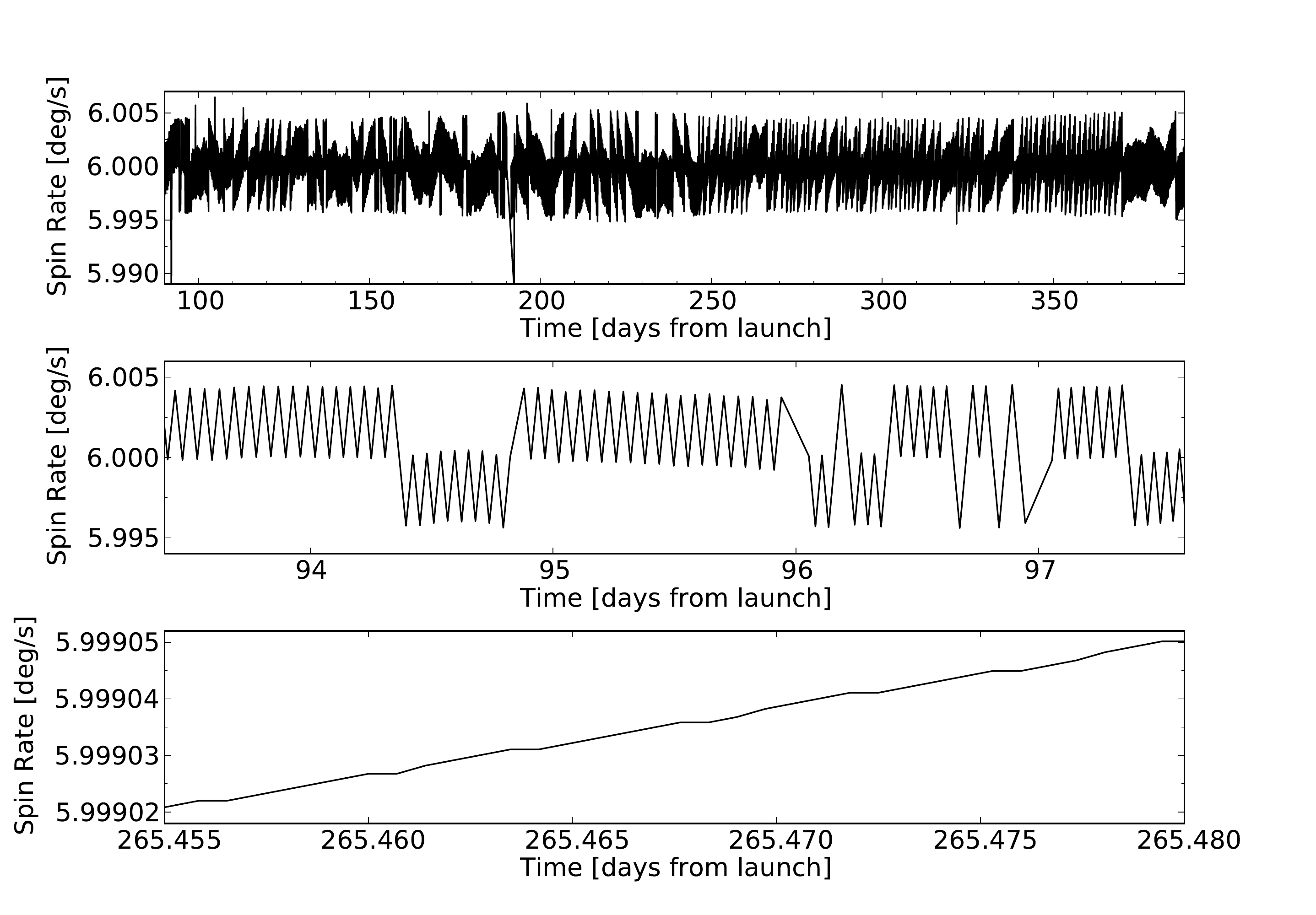} \\
   \caption{The top panel shows the measured spin rate over the
   coverage period, which is always within $\pm$0.1\%\ of $1\,$rpm. The
   spin rate typically adopts one of three values as can be seen
   in the middle panel, which is a zoom into a period of several
   days. In the bottom panel we show the typical drift of the spin rate
   within a dwell, due to small residual torques applied to the satellite. 
   }
              \label{FigSpinRate}
    \end{figure}

\Planck\ rotates about its principal axis of inertia at $1\,$rpm, with a
precision of $\pm$0.1\% (see Fig.~\ref{FigSpinRate}). The observed variation of
the spin rate is very systematic due to the following operational features.
\begin{itemize}
  \item The thruster used for a manoeuvre is selected depending
   on whether the spin rate preceding the manoeuvre is below or above its
   nominal value. Each thruster has a slightly different ``minimum
   thrust level,'' which determines the  spin rate actually achieved. Therefore,
   the spin rate after each manoeuvre will toggle between two different
   spin rate states which bound the nominal value.
   If one of these states is very close to the nominal value, drift
   during the dwell period (see next item) could cause it to change
   from one side to the other of the nominal value, and thus to toggle
   on the next manoeuvre to a ``third'' spin rate state.
  \item  Within a dwell period, the spin rate drifts slightly (typically
   $10^{-6}\,{\rm deg}\,{\rm s}^{-1}$ per minute) due to
  residual torques on the satellite, caused by solar radiation pressure and
  exhaust of helium from the dilution cooler system.
\end{itemize}

\begin{figure}
   \centering
   \includegraphics[width=0.5\textwidth]{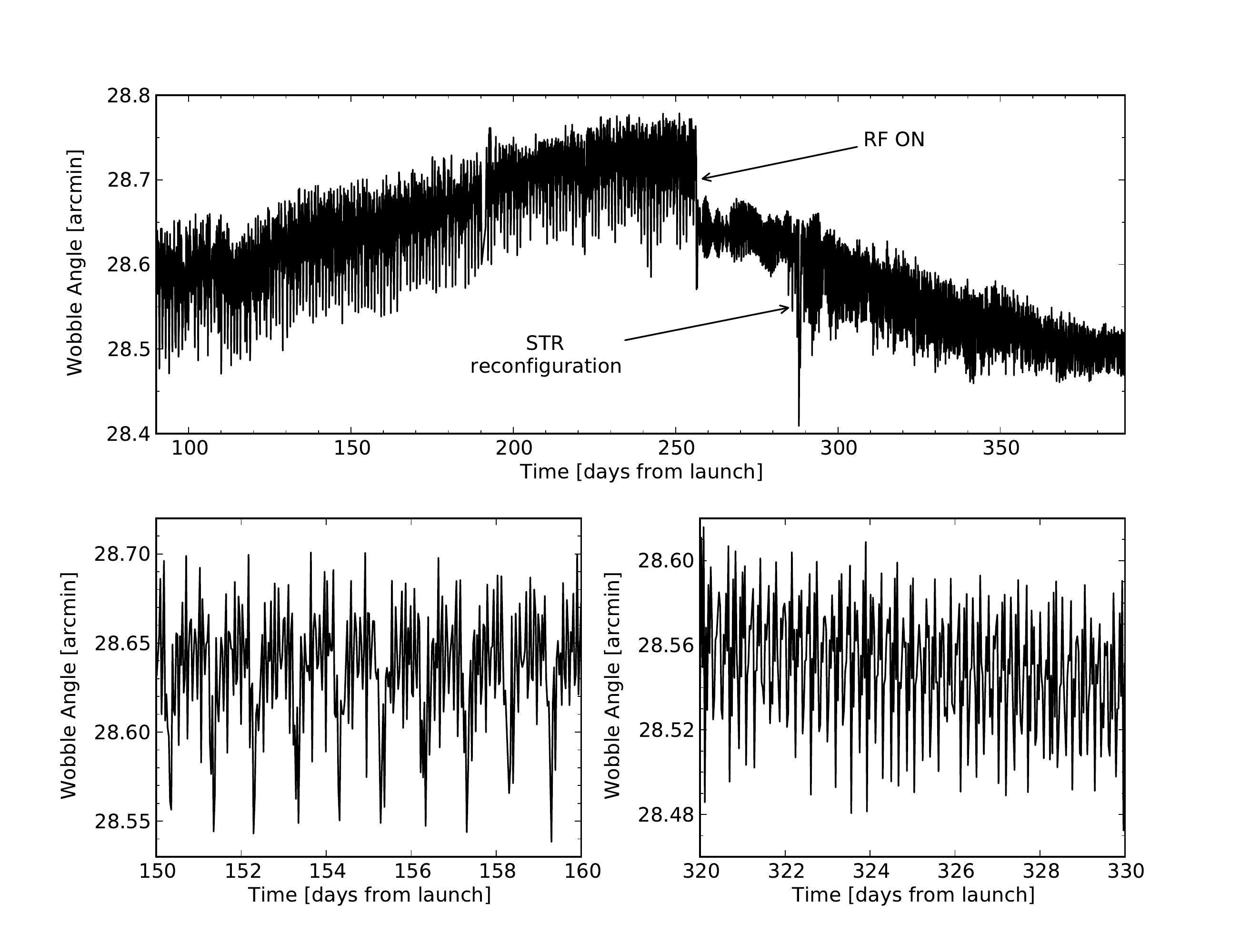} \\
   \caption{Top: evolution of the measured offset between the principal axis of
   inertia and the geometric axis of the satellite (one point is plotted for each
   dwell period of approximately 50 minutes). The time variations observed
   are not real changes in the offset angle, but are instead due to
   thermoelastic deformations in the panel that supports the star
   trackers. Bottom: zooming in on short periods of time before (left) and after (right) the
   RF transponder was turned permanently on, reveals periodic
   variations (before, dominated by the daily RF On-Off switching; after,
   dominated by thermal
   control cycling of nearby units with typically one hour periodicity).
   }
              \label{FigWobble}
    \end{figure}

The principal axis of inertia, about which \Planck\ rotates, is offset from
the geometric axis by $\sim28.6$\arcmin\
(see Fig.~\ref{FigWobble}). The time evolution of the
measured offset angle shows a long-timescale variation which is
clearly linked to the seasonal power input variations on the solar
array; this effect is not a real variation of the offset angle but
is instead due to a thermoelastic deformation of the SVM
panel that holds the star trackers\footnote{However, it appears as real in
the filtered attitude.}. Other thermoelastic deformations
that give rise to similar effects are related to specific operations
which have a thermal impact (see Fig.~\ref{FigWobble}).  The
dominant (false) offset angle variation before 25 January 2010 is
due to the daily thermal impact of the RF transponder being switched
on and off, and after that date it is related to thermal control
cycles in electronic units located near the star trackers; the
peak-to-peak amplitude of the effect is of order 0.15\arcmin\
before and 0.08\arcmin\ after 25 January. These effects can be
correlated to the temperature of the units responsible. 
They easily mask the real variation of the offset angle, which is due to
gradual depletion of the fuel and Helium tanks, and is of order
2.5\arcsec\ per month\footnote{The variation is
approximately 5\arcmin\ per $50\,$kg of fuel expended.
$\sim170\,{\rm g}\,{\rm month}^{-1}$ are expended in scanning 
manoeuvres and $\sim50\,$g in each orbital maintenance manoeuvre, currently
performed once every 8 weeks. Approximately $215\,{\rm g}\,{\rm month}^{-1}$
of Helium are vented to space by the HFI dilution cooler.}.  This
real variation of the wobble causes a corresponding change over time
of the radius of the circle which each detector traces on the sky.

\begin{figure*}
   \centering
   \includegraphics[width=0.9\textwidth]{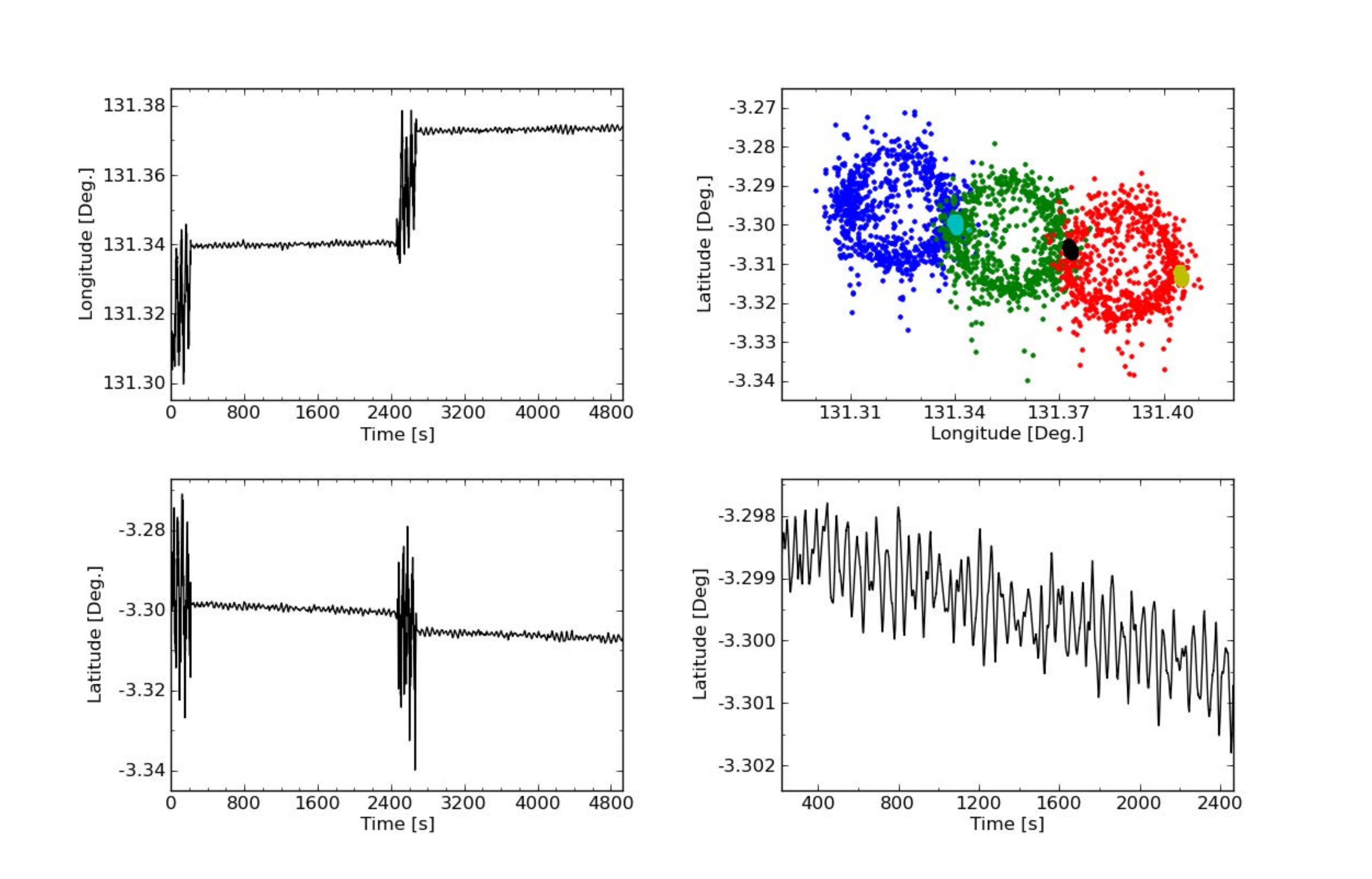} \\
   \caption{
    Left panels: motion of the spin axis (i.e., the direction of the principal
    axis of inertia) in latitude (top) and longitude
    (bottom) as a function of time, over a typical period of 70 minutes, 
    including two manoeuvres clearly identified by the large excursions in
    latitude and longitude. The dwell periods show the much smaller amplitude
    motions due to nutation and drift arising from solar radiation pressure.
    Top right: motion of the spin axis in Galactic
    latitude and longitude for a sequence of three manoeuvres and
    dwells. The attitude measurements during the slew are not
    filtered and are therefore very noisy; the measurements during each
    dwell are strongly clustered and show up as dense spots.
    Bottom right: zoom on a portion (about 30 minutes) of a
    dwell showing clearly the periodic motions which are a combination of
    the
    drift and residual nutation (period $\sim$5.4 minutes).
  }
              \label{FigTypManDwell}
    \end{figure*}

As described in \cite{tauber2010a}, \Planck's spin axis is displaced by
2\arcmin\ approximately every 50 minutes. A typical sequence of
manoeuvres and dwells is illustrated in Fig.~\ref{FigTypManDwell}.
Each manoeuvre is carried out as a sequence of three thrusts spaced
over three minutes (1st impulse -- two minutes wait -- 2nd impulse --
one minute wait -- 3rd impulse), designed to cancel nutation as much as
possible. Each manoeuvre lasts an average of about 220 seconds, as defined
by on-board software mode transitions\footnote{The ``start'' of the manoeuvre
mode is defined when the first thrust command
is issued, triggering the actual thrust up to a half a minute later,
and the ``end'' takes place immediately
after the last thrust in the manoeuvre sequence.},
which are used on the ground to trigger the end and start times of 
attitude filtering. 

\begin{figure}
   \centering
   \includegraphics[width=0.45\textwidth]{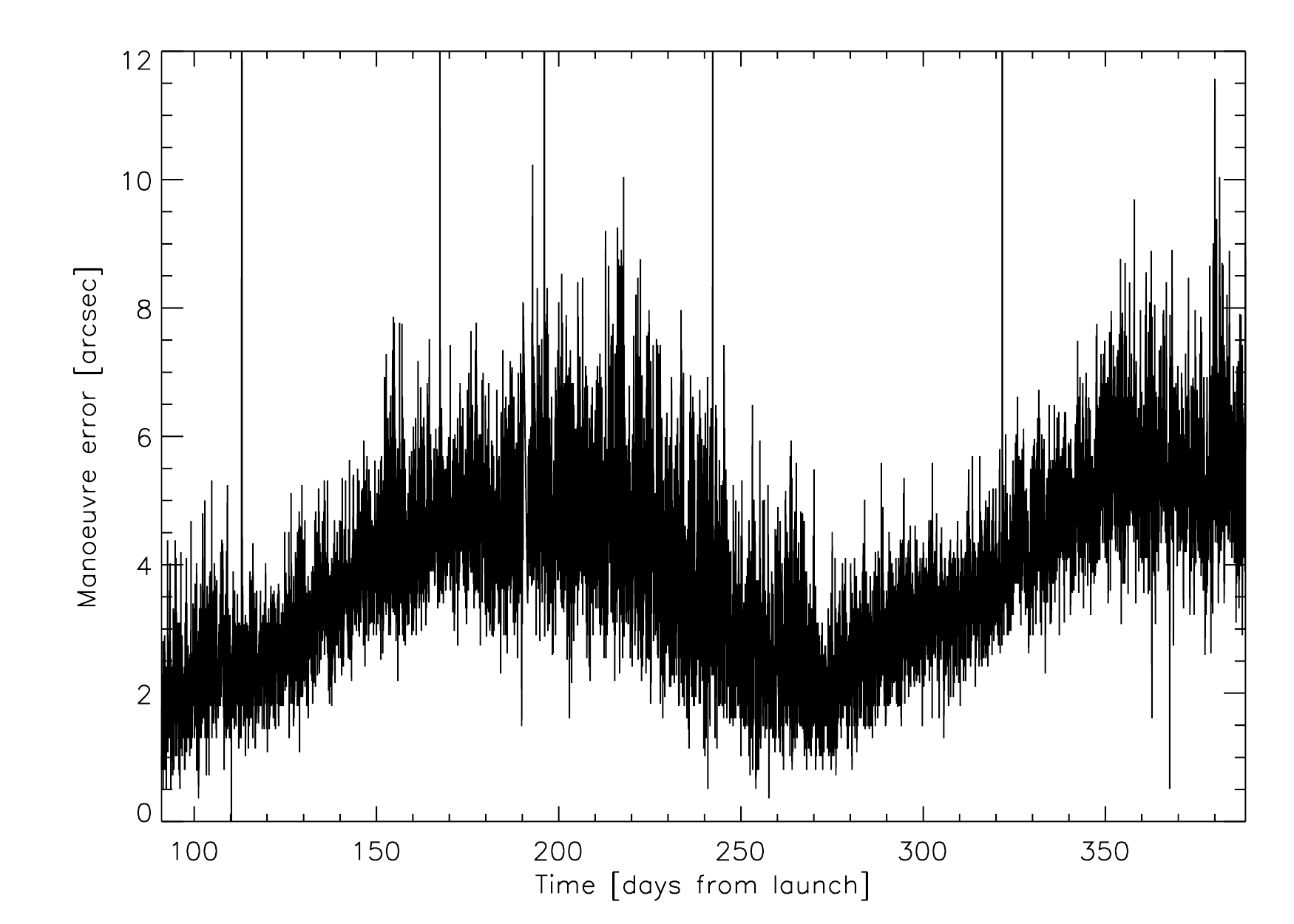} \\
   \caption{The plot shows the difference between the achieved and commanded pointing after
   each 2\arcmin\ manoeuvre; the difference is correlated with the duration of the
   previous dwell (see text). The vertical lines correspond to
   the very few occasions when the manoeuvre sequence did not execute as planned,
   resulting in anomalously high pointing errors.
   }
              \label{FigManError}
    \end{figure}

The pointing achieved after each reorientation is of course not
exactly the commanded one; the difference, which varies between 2
and 8\arcsec\ during the coverage period, is shown in
Fig.~\ref{FigManError}. This variation is systematically related to the
duration of the dwell preceding the manoeuvre, since the preceding
pointing drifts due to solar radiation pressure, and the angular
amplitude of the following manoeuvre changes correspondingly. The
thrust sequence required for each manoeuvre is computed on-board
based on the known mass properties of the satellite and known
thruster response functions; the error made on each manoeuvre is
mainly driven by the (imperfect) on-board knowledge of these
properties, and therefore depends systematically on the amplitude of
the manoeuvre. On very few occasions (visible in Fig.~\ref{FigManError}),
the thruster sequence performance did not execute as planned, and
resulted in a much larger manoeuvre error.

\begin{figure}
   \centering
   \includegraphics[width=0.5\textwidth]{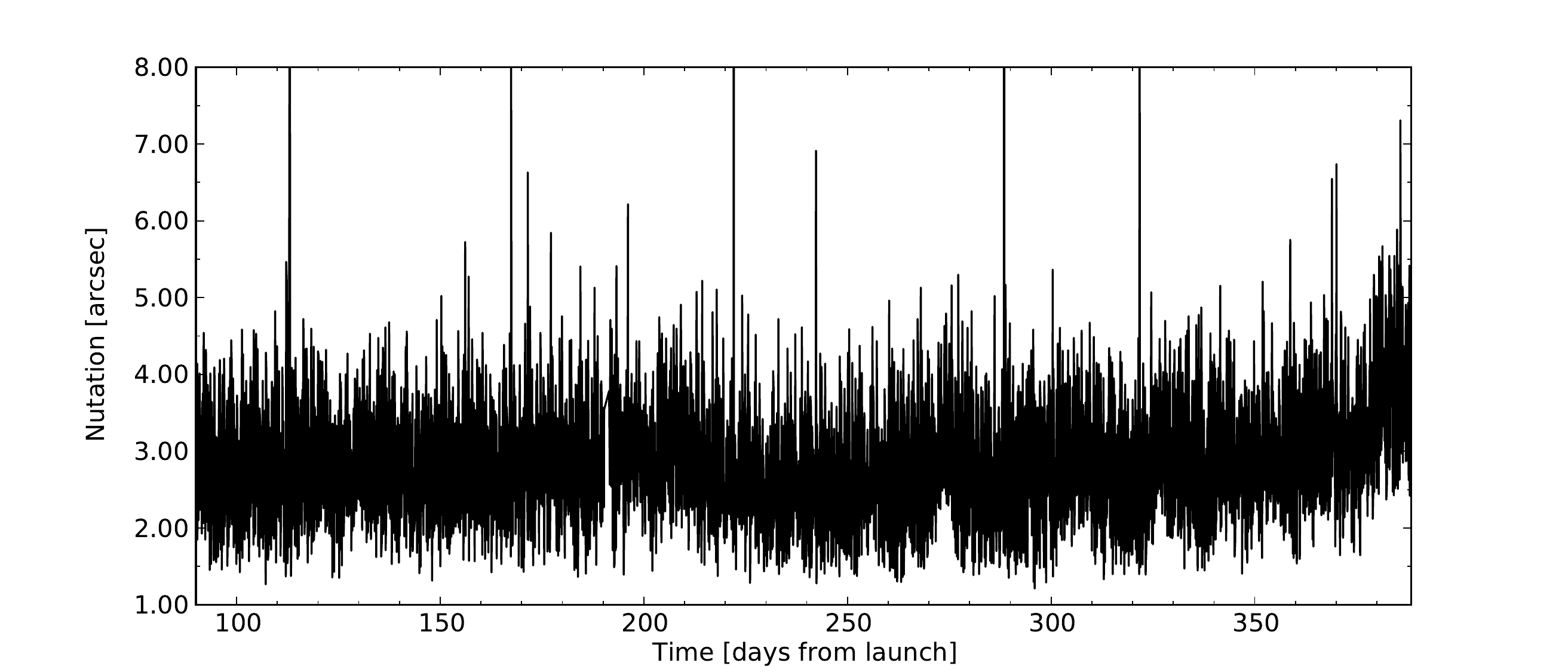} \\
   \caption{The peak amplitude of the residual nutation averaged within each
   dwell time, over the coverage period.
   }
              \label{FigNutation}
    \end{figure}

Although the thruster sequence is designed to damp nutation, it does
not do so perfectly. The peak amplitude of the residual nutation is
typically 3\arcsec\ and does not vary significantly in time (see
Fig.~\ref{FigNutation}). The period of the nutation is $5.425\pm0.010$
minutes, determined by the inertial properties of the satellite.
Neither the amplitude nor the period
of the nutation are observed to drift during a dwell period.

\begin{figure}
   \centering
   \includegraphics[width=0.45\textwidth]{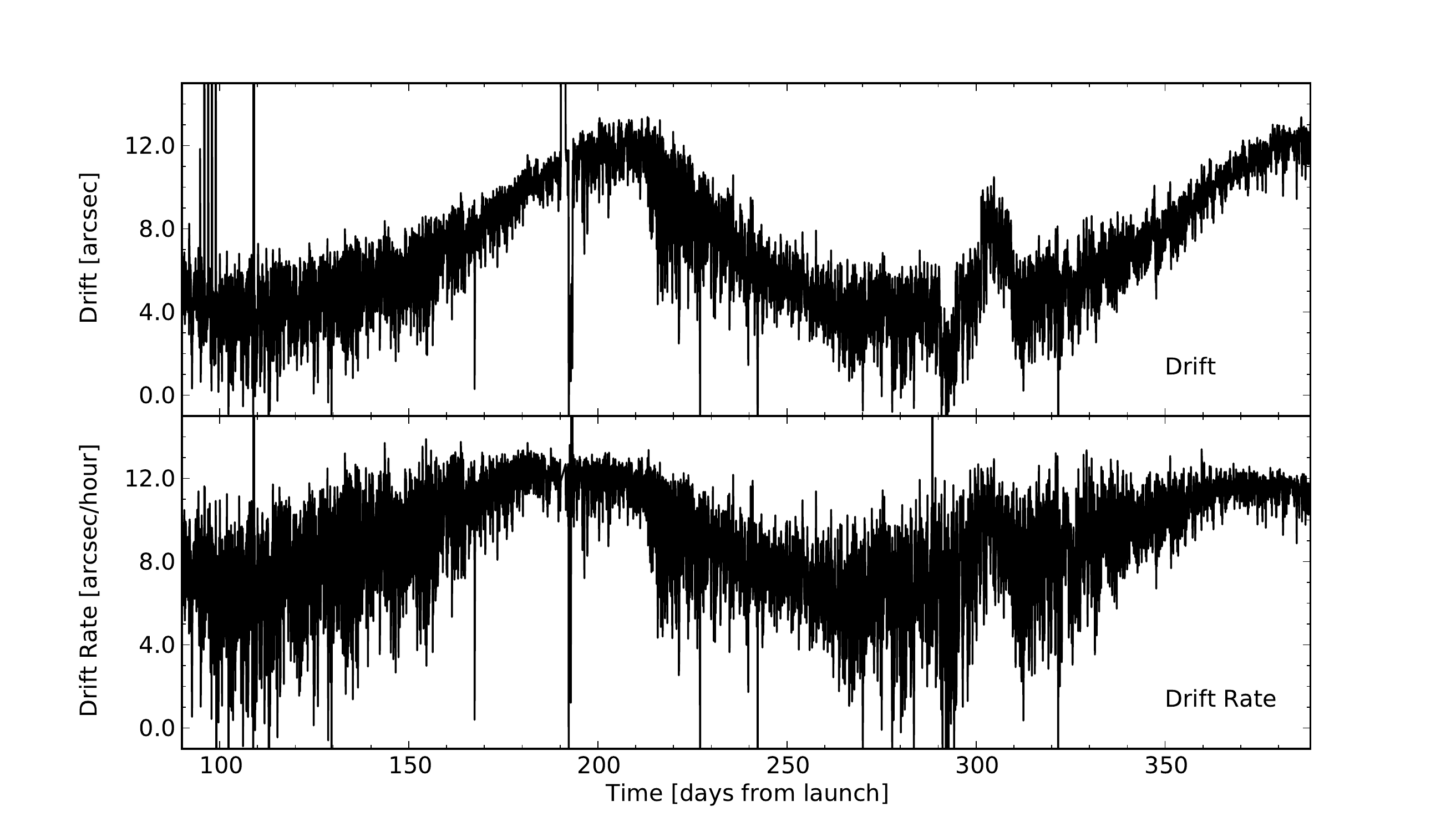} \\
   \caption{The total drift due to solar radiation pressure
   within each dwell, which depends on the time length of the dwell
   (see Fig.~\ref{FigScanning}). The drift rate (shown in the lower panel) is approximately constant at
   $\sim$9\arcsec${\rm hr}^{-1}$.
   }
              \label{FigDrift}
    \end{figure}

The last major characteristic of pointing within dwells is the drift
due to solar radiation pressure. The total amplitude of the drift
within each dwell varies between $\sim$5 and $\sim$12\arcsec,
depending on the duration of each dwell (see Fig.~\ref{FigDrift}).
The rate of drift varies between $\sim$4 and
$\sim$10\arcsec\ hour$^{-1}$,  weakly correlated with the Solar Aspect
Angle (which varies by very small amounts throughout the
mission). 

The pointing characteristics described above are summarised in
Table~\ref{TabPointing}.

\begin{table}
\caption{\Planck\ Pointing Performance}             
\label{TabPointing}      
\vskip -18pt
\setbox\tablebox=\vbox{
\newdimen\digitwidth
\setbox0=\hbox{\rm 0}
\digitwidth=\wd0
\catcode`*=\active
\def*{\kern\digitwidth}
\newdimen\signwidth
\setbox0=\hbox{+}
\signwidth=\wd0
\catcode`!=\active
\def!{\kern\signwidth}
\newdimen\pointwidth
\setbox0=\hbox{.}
\pointwidth=\wd0
\catcode`@=\active
\def@{\kern\pointwidth}
\halign{#\hfil\tabskip 1.0em&
\hfil#\hfil&
\hfil#\hfil&
\hfil#\hfil\tabskip 8pt\cr
\noalign{\vskip 3pt\hrule\vskip 1.5pt\hrule\vskip 5pt}
&  Median  &  Std. dev. & unit \cr
\noalign{\vskip 4pt\hrule\vskip 6pt}
Spin rate                           & 6.0001  & 0.0027 & ${\rm deg}\,{\rm s}^{-1}$ \cr
Small manoeuvre accuracy            & 3.6****  & 1.8*** & arcsec  \cr
Residual nutation amplitude & & & \cr
\noalign{\vskip -3pt}
\ after manoeuvre                   & 2.68***  & 1.75** & arcsec \cr
Drift rate during inertial pointing & 9.14***  & 3.0*** &
 ${\rm arcsec}\,{\rm hr}^{-1}$ \cr
\noalign{\vskip 3pt\hrule\vskip 4pt}
} }
\endPlancktable
\end{table}

\section{Payload Performance}\label{sec:PayloadPerformance}

The performance of the payload (i.e., two instruments and telescope) is 
described in detail in \citep[][LFI]{planck2011-1.4}
and \citep[][HFI]{planck2011-1.5}, and 
summarised in Table~\ref{table:Instrument_performance}. We note that:
\begin{itemize}
\item the angular resolution measured on planets is within a few per cent of that
predicted on the ground \citep{tauber2010b}
\item the instantaneous sensitivity
 of the \Planck\ LFI \citep{planck2011-1.4} and HFI \citep{planck2011-1.5} channels
 is estimated to be approximately 10\%\ larger than that measured on the ground 
 and extrapolated to launch conditions \citep{tauber2010a}. 
 For LFI the excess
 is understood to be due to inaccuracy in the calibration constants as
 measured in ground tests \citep{planck2011-1.4}; for HFI it is
 expected to be due to systematic effects remaining in the data at the
 current level of processing \citep{planck2011-1.5}.
\item the photometric calibration uncertainty quoted is conservatively
 based on the current knowledge of systematic effects and data processing
 pipelines \citep{planck2011-1.6, planck2011-1.7}.
 There is no reason to believe that the mission
 goals (1\%\ in CMB channels and 3\%\ at the highest frequencies) will not be 
 reached for all \Planck\ channels in due time. 
\item the point source sensitivities quoted correspond to the fluxes of the
 faintest sources included in the ERCSC (\cite{planck2011-1.10}). Since the
 ERCSC is a high-reliability catalogue, based on very robust
 extraction from only the first all-sky survey, these levels will certainly
 improve substantially in the legacy catalogues which will be delivered in
 January 2013.
\end{itemize}

Table~\ref{table:Instrument_performance} confirms the findings of the early
CPV activities, namely that the basic performance parameters of the scientific
payload of \Planck\ are very close to what was expected based on
the measurements made on the ground before launch.

\begin{table*}[tmb]
\begingroup
\newdimen\tblskip \tblskip=5pt
\caption{\Planck\ performance parameters determined from flight data.} 
                            \label{table:Instrument_performance}
\nointerlineskip
\vskip -3mm
\footnotesize
\setbox\tablebox=\vbox{
    \newdimen\digitwidth
    \setbox0=\hbox{\rm 0}
    \digitwidth=\wd0
    \catcode`*=\active
    \def*{\kern\digitwidth}
    \newdimen\signwidth
    \setbox0=\hbox{+}
    \signwidth=\wd0
    \catcode`!=\active
    \def!{\kern\signwidth}
\def\muKCMBs{\ifmmode \,\mu$K$_{\rm CMB}$\,s$^{1/2}\else \,$\mu$K$_{\rm CMB}$\,s$^{1/2}$\fi}
\halign{\hbox to 1.05in{#\leaderfil}\tabskip=1.0em&
      \hfil#\hfil\tabskip=1.0em&
      \hfil#\hfil\tabskip=1.5em
&
      \hfil#\hfil\tabskip=0.5em&
      \hfil#\hfil\tabskip=1.5em&
      \hfil#\hfil\tabskip=0.1em&
      \hfil#\hfil\tabskip=0.2em&
      \hfil#\hfil\tabskip=0.5em&
      \hfil#\hfil\tabskip=0pt\cr
\noalign{\doubleline}
\omit&&&&&\multispan2\hfil W{\sc hite-noise}$^{\rm d}$\hfil \cr
\omit&&&\multispan2\hfil M{\sc ean} B{\sc eam}$^{\rm c}$\hfil&\multispan2\hfil S{\sc ensitivity}\hfil&C{\sc alibration}$^{\rm e}$&F{\sc aintest} S{\sc ource}$^{\rm f}$\cr
\omit&&$\nu_{\rm center}$$^{\rm b}$&\multispan2\hrulefill&\multispan2\hrulefill&U{\sc ncertainty}&{\sc in} ERCSC $|b|>30\deg$\cr
\omit\hfil C{\sc hannel}\hfil&$N_{\rm detectors}$$^{\rm a}$&[GHz]&FWHM&Ellipticity&[\muKRJs]&[\muKCMBs]&[\%]&[mJy]\cr
\noalign{\vskip 3pt\hrule\vskip 5pt}
*30\,GHz&*4&**\phantom{.}\getsymbol{LFI:center:frequency:30GHz}&\getsymbol{LFI:FWHM:30GHz}& 1.38&\getsymbol{LFI:white:noise:sensitivity:30GHz}&146.8&1&480\cr
*44\,GHz&*6&**\phantom{.}\getsymbol{LFI:center:frequency:44GHz}&\getsymbol{LFI:FWHM:44GHz}& 1.26&\getsymbol{LFI:white:noise:sensitivity:44GHz}&173.1&1&585\cr
*70\,GHz&12&**\phantom{.}\getsymbol{LFI:center:frequency:70GHz}&\getsymbol{LFI:FWHM:70GHz}& 1.27&\getsymbol{LFI:white:noise:sensitivity:70GHz}&152.6&1&481\cr
100\,GHz&*8&\getsymbol{HFI:center:frequency:100GHz}& *\getsymbol{HFI:FWHM:Mars:100GHz}&\getsymbol{HFI:beam:ellipticity:Mars:100GHz}&*17.3&*22.6&2&344\cr
143\,GHz&11&\getsymbol{HFI:center:frequency:143GHz}& *\getsymbol{HFI:FWHM:Mars:143GHz}&\getsymbol{HFI:beam:ellipticity:Mars:143GHz}&**8.6&*14.5&2&206\cr
217\,GHz&12&\getsymbol{HFI:center:frequency:217GHz}& *\getsymbol{HFI:FWHM:Mars:217GHz}&\getsymbol{HFI:beam:ellipticity:Mars:217GHz}&**6.8&*20.6&2&183\cr
353\,GHz&12&\getsymbol{HFI:center:frequency:353GHz}& *\getsymbol{HFI:FWHM:Mars:353GHz}&\getsymbol{HFI:beam:ellipticity:Mars:353GHz}&**5.5&*77.3&2&198\cr
545\,GHz&*3&\getsymbol{HFI:center:frequency:545GHz}& *\getsymbol{HFI:FWHM:Mars:545GHz}&\getsymbol{HFI:beam:ellipticity:Mars:545GHz}&**4.9&\dots&7&381\cr
857\,GHz&*3&\getsymbol{HFI:center:frequency:857GHz}& *\getsymbol{HFI:FWHM:Mars:857GHz}&\getsymbol{HFI:beam:ellipticity:Mars:857GHz}&**2.1&\dots&7&655\cr
\noalign{\vskip 5pt\hrule\vskip 3pt}}}
\endPlancktablewide
\tablenote a For 30, 44, and 70\,GHz, each ``detector'' is a linearly polarised radiometer.  There are two (orthogonally polarized) radiometers behind each horn.  Each radiometer has two diodes, both switched at high frequency between the sky and a blackbody load at $\sim4$\,K \citep{planck2011-1.4}. For 100\,GHz and above, each ``detector'' is a bolometer \citep{planck2011-1.5}. Most of the bolometers are sensitive to polarisation, in which case there are two orthogonally polarised detectors behind each horn; some of the detectors are spider-web bolometers (one per horn) sensitive to the total incident power.\par
\tablenote b Mean center frequency of the $N$ detectors at each frequency.\par
\tablenote c Mean optical properties of the $N$ beams at each frequency; FWHM $\equiv$ FWHM of circular Gaussian with the same volume.  
Ellipticity gives the ratio of major axis to minor axis for a best-fit elliptical Gaussian. 
In the case of HFI, the mean values quoted are the result of 
averaging the values of total-power and polarisation-sensitive bolometers, weighted by the number of channels and after removal of
those affected by random telegraphic noise.  The actual point spread function of an unresolved object on the sky depends not only on the optical properties of the beam, but also on sampling and time domain filtering in signal processing, and the way the sky is scanned.  For details on these aspects see \S\,4 of \cite{planck2011-1.4}, \S\,4 of \cite{planck2011-1.6}, \S\,4.2 of \cite{planck2011-1.5}, and \S\,6.2 of \cite{planck2011-1.7}.\par
\tablenote d Uncorrelated noise on the sky in 1\,s for the array of $N$ detectors, in Rayleigh-Jeans units and in thermodynamic CMB units.  For a preliminary discussion of correlated noise and systematic effects, see \cite{planck2011-1.4}, \cite{planck2011-1.5} , \cite{planck2011-1.6} , and \cite{planck2011-1.7} .\par
\tablenote e  Absolute uncertainty, based on the known amplitude of the CMB dipole up to 353\,GHz, and on FIRAS at 545 and 857\,GHz \citep{planck2011-1.6,planck2011-1.7}.\par
\tablenote f Flux density of the faintest source included in the ERCSC \citep{planck2011-1.10}.\par
\endgroup
\end{table*}

The ultimate performance of \Planck\ also depends on its operational lifetime. 
\Planck\ is a cryogenic mission \citep{planck2011-1.3}, whose nominal
lifetime in routine operations 
(i.e., excluding transfer to orbit, commissioning and performance verification
phases) was initially set to 15 months, allowing it to complete two full
surveys of the sky within that period.  
Its actual lifetime is limited by the active coolers required to operate the
\Planck\ detectors, as listed hereafter.  
\begin{itemize}
\item A $^3$He-$^4$He dilution refrigerator, which cools the HFI bolometers to
$0.1\,$K.  The $^3$He and $^4$He gas are stored in tanks and vented to space
after the dilution process.  
In-flight measurements of tank depletion predict that the $^3$He gas will run
out at the end of January 2012 \citep{planck2011-1.3}.
\item	A hydrogen sorption refrigerator, which cools the LFI radiometers to
 $20\,$K and provides a first pre-cooling stage for the HFI bolometer system.
 Its lifetime is limited by gradual degradation of the sorbent material.
 Two units fly on Planck: the first has provided cooling until August 2010; 
 the second came into operation thereafter and is currently predicted to
 allow operation until December 2011 \citep{planck2011-1.3}. 
 A further increase of lifetime may be obtained by applying on board the
 process of ``regeneration'' to the material.
\end{itemize}

Overall, the cooling system lifetime is at least one year above the 
nominal mission span, and no other spacecraft or payload factors impose
additional limitations. Therefore, barring unexpected
failures, \Planck\ will continue surveying the sky until at least the end of 2011.

\section{Conclusions}
This paper summarises the performance of the \Planck\ satellite during
its first year of survey operations, in the areas most relevant for
scientific analysis of the \Planck\ data. Detailed descriptions of
all aspects of the payload are provided in accompanying papers in this
issue. It can be concluded that the major elements of the
satellite's performance exceed their original technical
requirements, and the scientific performance approaches the mission goals.

After an astoundingly smooth first year of survey operations,
\Planck\ continues to observe the sky and gather high quality scientific
data, and is expected to do so as
long as the cryogenic chain can keep the $100\,$mK stage near its
nominal temperature, i.e., to the end of 2011 or possibly early 2012.
Following the end of mission operations, the next major milestone in the project 
will be the release of the first set of timeline and map data products,
currently foreseen in January 2013.


\begin{acknowledgements}

\Planck\ is too large a project to allow full acknowledgement of
all contributions by individuals, institutions, industries, and
funding agencies. The main entities involved in the mission operations are as
follows. 
The European Space Agency operates the satellite via its Mission
Operations Centre located at ESOC (Darmstadt, Germany) and coordinates
scientific operations via the \Planck\ Science Office located at ESAC (Madrid, Spain).
Two Consortia, comprising around 100
scientific institutes within Europe, the USA, and Canada, and funded by
agencies from the participating countries, developed the
scientific instruments LFI and HFI, and 
continue to operate them via Instrument Operations Teams located in Trieste
(Italy) and Orsay (France).
The Consortia are also responsible for scientific
processing of the
acquired data. The Consortia are led by the Principal Investigators:
J.-L. Puget in France for HFI (funded principally by CNES and CNRS/INSU-IN2P3)
and N.  Mandolesi in Italy for LFI (funded principally via ASI). 
NASA's US \Planck\ Project, based at JPL and involving
scientists at many US institutions, contributes 
significantly to the efforts of these two Consortia. 
A third Consortium, led by H. U. Norgaard-Nielsen and supported by the 
Danish Natural Research Council, contributed to the reflector programme.
The author list
for this paper has been selected by the \Planck\ Science Team from the Planck Collaboration, and
is composed of individuals from all of the above entities who have
made multi-year contributions to the development of the mission. It
does not pretend to be inclusive of all contributions. 
A description of the Planck Collaboration and a list of its members, 
indicating which technical or scientific activities they have been involved in, 
can be found at \emph{(http://www.rssd.esa.int/index.php?project=PLANCK\& page=Planck\_ Collaboration)}. 
The Planck Collaboration acknowledges the support of: ESA; CNES and CNRS/INSU-IN2P3-INP (France); 
ASI, CNR, and INAF (Italy); NASA and DoE (USA); STFC and UKSA (UK); 
CSIC, MICINN and JA (Spain); Tekes, AoF and CSC (Finland); DLR and MPG (Germany); 
CSA (Canada); DTU Space (Denmark); SER/SSO (Switzerland); RCN (Norway); SFI (Ireland); 
FCT/MCTES (Portugal); and DEISA (EU).

\end{acknowledgements}

\bibliographystyle{aa}

\bibliography{16464_Planck_bib}

\end{document}